\documentclass[twocolumn]{bmcart}% uncomment this for twocolumn layout and comment line below
\usepackage{amsthm,amsmath}
\usepackage{tabto}
\usepackage[utf8]{inputenc} %unicode support
\usepackage{url}
\def\includegraphics{}

\usepackage{graphicx}
\startlocaldefs
\endlocaldefs

\begin{document}

%%% Start of article front matter
\begin{frontmatter}

\begin{fmbox}
\dochead{Research}


\title{Generating Chord Progression from Melody with Flexible Harmonic Rhythm and Controllable Harmonic Density}

%%%%%%%%%%%%%%%%%%%%%%%%%%%%%%%%%%%%%%%%%%%%%%
%%                                          %%
%% Enter the authors here                   %%
%%                                          %%
%% Specify information, if available,       %%
%% in the form:                             %%
%%   <key>={<id1>,<id2>}                    %%
%%   <key>=                                 %%
%% Comment or delete the keys which are     %%
%% not used. Repeat \author command as much %%
%% as required.                             %%
%%                                          %%
%%%%%%%%%%%%%%%%%%%%%%%%%%%%%%%%%%%%%%%%%%%%%%

\author[
addressref={aff1},
email={shangda@mail.ccom.edu.cn},
noteref={n1}
]{\inits{S.}\fnm{Shangda} \snm{Wu}}
\author[
addressref={aff1},
email={yewyang@mail.ccom.edu.cn},
noteref={n1}
]{\inits{Y.}\fnm{Yue} \snm{Yang}}
\author[
addressref={aff1},
email={wzw@mail.ccom.edu.cn},
noteref={n1}
]{\inits{Z.}\fnm{Zhaowen} \snm{Wang}}
\author[
addressref={aff1},
email={lxiaobing@ccom.edu.cn}
]{\inits{X.}\fnm{Xiaobing} \snm{Li}}
\author[
addressref={aff1,aff2},
email={sms@tsinghua.edu.cn},
corref={aff1,aff2}
]{\inits{M.}\fnm{Maosong} \snm{Sun}}

%%%%%%%%%%%%%%%%%%%%%%%%%%%%%%%%%%%%%%%%%%%%%%
%%                                          %%
%% Enter the authors' addresses here        %%
%%                                          %%
%% Repeat \address commands as much as      %%
%% required.                                %%
%%                                          %%
%%%%%%%%%%%%%%%%%%%%%%%%%%%%%%%%%%%%%%%%%%%%%%

\address[id=aff1]{%                           % unique id
  \orgdiv{Department of Music AI and Information Technology},             % department, if any
  \orgname{Central Conservatory of Music},          % university, etc
  \city{Beijing},                              % city
  \cny{China}                                    % country
}
\address[id=aff2]{%
  \orgdiv{Department of Computer Science and Technology},
  \orgname{Tsinghua University},          % university, etc
  \city{Beijing},                              % city
  \cny{China}                                    % country
}

%%%%%%%%%%%%%%%%%%%%%%%%%%%%%%%%%%%%%%%%%%%%%%
%%                                          %%
%% Enter short notes here                   %%
%%                                          %%
%% Short notes will be after addresses      %%
%% on first page.                           %%
%%                                          %%
%%%%%%%%%%%%%%%%%%%%%%%%%%%%%%%%%%%%%%%%%%%%%%

\begin{artnotes}
%\note{Sample of title note}     % note to the article
\note[id=n1]{Equal contribution} % note, connected to author
\end{artnotes}

%\end{fmbox}% comment this for two column layout

%%%%%%%%%%%%%%%%%%%%%%%%%%%%%%%%%%%%%%%%%%%%%%%
%%                                           %%
%% The Abstract begins here                  %%
%%                                           %%
%% Please refer to the Instructions for      %%
%% authors on https://www.biomedcentral.com/ %%
%% and include the section headings          %%
%% accordingly for your article type.        %%
%%                                           %%
%%%%%%%%%%%%%%%%%%%%%%%%%%%%%%%%%%%%%%%%%%%%%%%

\begin{abstractbox}

\begin{abstract} % abstract
Melody harmonization, which involves generating a chord progression that complements a user-provided melody, continues to pose a significant challenge. A chord progression must not only be in harmony with the melody, but also interdependent on its rhythmic pattern. While previous neural network-based systems have been successful in producing chord progressions for given melodies, they have not adequately addressed controllable melody harmonization, nor have they focused on generating harmonic rhythms with flexibility in the rates or patterns of chord changes. This paper presents AutoHarmonizer, a novel system for harmonic density-controllable melody harmonization with such a flexible harmonic rhythm. AutoHarmonizer is equipped with an extensive vocabulary of 1,462 chord types and can generate chord progressions that vary in harmonic density for a given melody. Experimental results indicate that the AutoHarmonizer-generated chord progressions exhibit a diverse range of harmonic rhythms and that the system's controllable harmonic density is effective.
\end{abstract}

%%%%%%%%%%%%%%%%%%%%%%%%%%%%%%%%%%%%%%%%%%%%%%
%%                                          %%
%% The keywords begin here                  %%
%%                                          %%
%% Put each keyword in separate \kwd{}.     %%
%%                                          %%
%%%%%%%%%%%%%%%%%%%%%%%%%%%%%%%%%%%%%%%%%%%%%%


% MSC classifications codes, if any
%\begin{keyword}[class=AMS]
%\kwd[Primary ]{}
%\kwd{}
%\kwd[; secondary ]{}
%\end{keyword}

\end{abstractbox}
\end{fmbox}% uncomment this for two column layout

\end{frontmatter}

%%%%%%%%%%%%%%%%%%%%%%%%%%%%%%%%%%%%%%%%%%%%%%%%
%%                                            %%
%% The Main Body begins here                  %%
%%                                            %%
%% Please refer to the instructions for       %%
%% authors on:                                %%
%% https://www.biomedcentral.com/getpublished %%
%% and include the section headings           %%
%% accordingly for your article type.         %%
%%                                            %%
%% See the Results and Discussion section     %%
%% for details on how to create sub-sections  %%
%%                                            %%
%% use \cite{...} to cite references          %%
%%  \cite{koon} and                           %%
%%  \cite{oreg,khar,zvai,xjon,schn,pond}      %%
%%                                            %%
%%%%%%%%%%%%%%%%%%%%%%%%%%%%%%%%%%%%%%%%%%%%%%%%

%%%%%%%%%%%%%%%%%%%%%%%%% start of article main body
% <put your article body there>

%%%%%%%%%%%%%%%%
%% Background %%
%%
\section*{Introduction}
In recent years, there has been considerable research effort devoted to developing the practical applications of neural networks in the field of music. Neural networks have enabled the implementation of automatic transcription \cite{DBLP:conf/mir/LiuZMHCZX21, DBLP:conf/ismir/Calvo-ZaragozaR18}, which involves converting audio signals into symbolic musical representations. Researchers also used neural networks to classify musical pieces by genre \cite{DBLP:conf/interspeech/GhosalK18,DBLP:conf/evoW/DervakosKS21} and even generate original music \cite{DBLP:journals/corr/abs-1709-01620,DBLP:conf/pimrc/CasiniMR18,DBLP:journals/csur/HerremansCC17}. This paper focuses on the task of melody harmonization, which involves creating a neural network-based system that generates chord progressions to accompany a given melody, with the added ability to control the harmonic rhythm.

In music, a chord is a combination of multiple notes that produce a harmonious sound. The transition of chords within a musical composition is known as harmonic rhythm or harmonic tempo. Melody harmonization systems, as studied in \cite{makris2016automatic, DBLP:conf/icmcs/SunWY22, DBLP:conf/ismir/TsushimaNIY17}, are designed to automatically generate suitable chords that accompany a given melody, essentially harmonizing the melody.

The process of harmonizing a melody involves selecting the appropriate chords that complement the melody's underlying tonality, structure, and rhythm. The harmonization system must analyze the melody's pitch, duration, and rhythmic patterns to identify the most appropriate chords to use at each point in the melody. The system's output should enhance the melody's expressiveness, while also maintaining a sense of coherence and musical logic.

Melody harmonization systems have a range of potential applications, including music composition, arranging, and production. These systems can be particularly useful for musicians who lack formal training in music theory or for those seeking inspiration for new musical compositions. Furthermore, these systems can facilitate the creation of harmonically complex and innovative musical arrangements by automating the process of writing chord progression.

\begin{figure}[t]
    \centering
        \begin{minipage}{8cm} 
            \includegraphics[width=\textwidth]{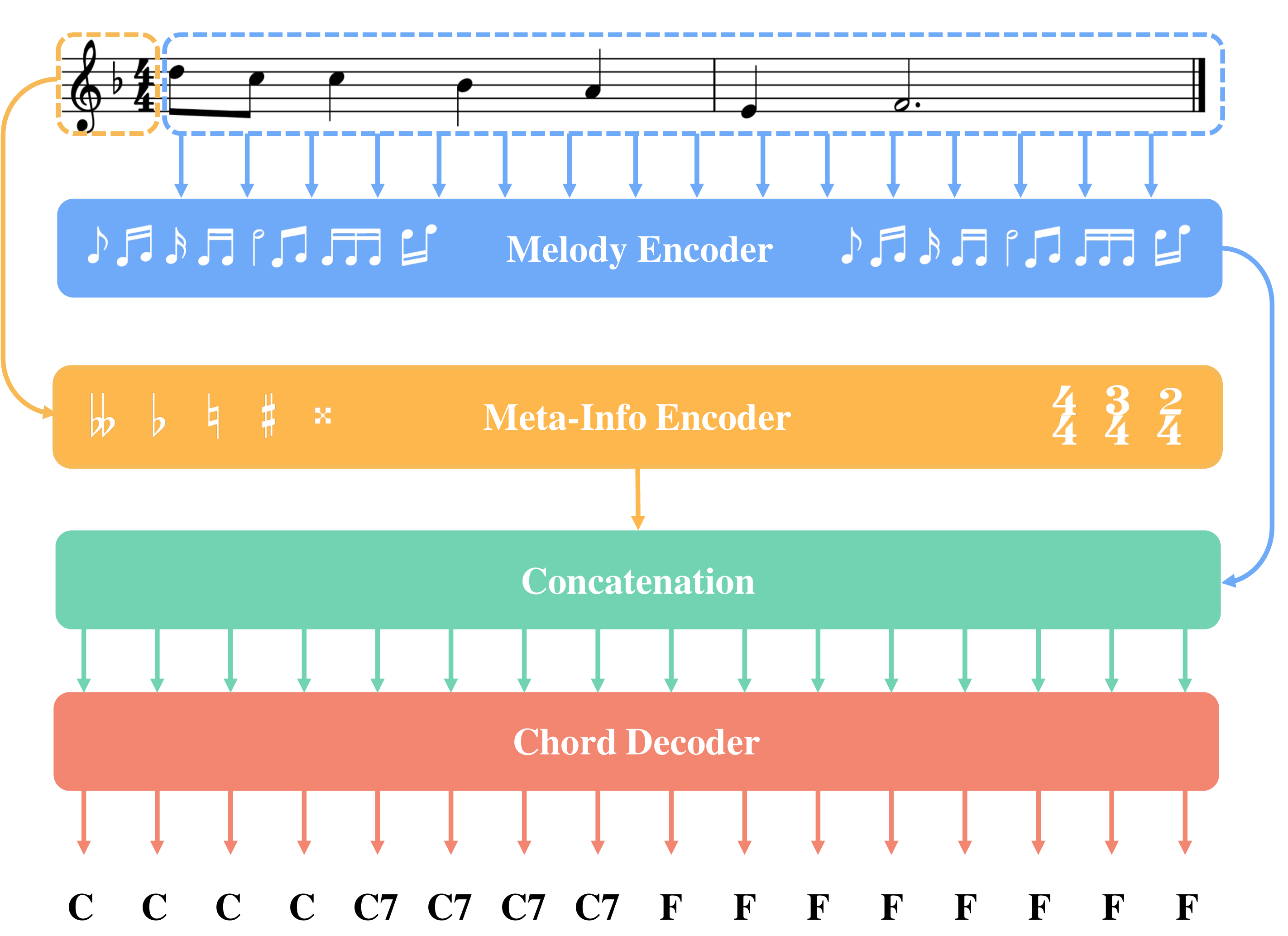}
        \end{minipage}
    \centering
    \caption{The architecture of AutoHarmonizer, which predicts chord symbols frame-by-frame (sixteenth note).}
\end{figure}

Melody harmonization is a complex and subjective task that has received considerable attention from researchers in recent years. However, existing works in this area \cite{DBLP:conf/ictai/BrunnerWWW17,DBLP:conf/icassp/SunCLCW21,DBLP:journals/corr/abs-2001-02360,DBLP:conf/ismir/ChenLCW21} have mainly focused on generating appropriate chords, while neglecting the equally important aspect of placing them within the proper musical context. Consequently, the aforementioned works suffer from limited flexibility in chord progression generation, as they tend to produce only a single chord for each bar or half-bar, leading to rigid harmonic rhythms that do not capture the subtleties of musical expression.

To address these challenges, it is necessary to develop more sophisticated models that can capture the full complexity of melody harmonization and generate musically satisfying and aesthetically pleasing results. Such models should take into account the wider musical context, including the relationships between different chords and their role in the overall harmonic structure. Furthermore, they should allow for greater flexibility in chord progression generation, allowing for variations in rhythmic and harmonic patterns that reflect the nuances of musical expression.

This study aims to develop a novel approach to achieve automatic melody harmonization with flexible harmonic rhythm, where chord progressions rhythmically match a given melody. To achieve this objective, the proposed approach, AutoHarmonizer, as shown in Fig. 1, generates chords on a sixteenth note basis (referred to as a `frame' in this context) instead of bar-by-bar. This modeling strategy better represents the task at hand and allows for more accurate harmonization. Additionally, time signatures are encoded to establish rhythmic relationships between melodies and chords. The controllable harmonic density, which refers to the degree of richness or sparsity in the generated chord progressions, based on the work of \cite{wu2022sampling}, has been implemented to enable customized harmonizations according to user preferences.

Contributions of this paper are summarized as follows:
\begin{itemize}
    \item The proposed model considers beat and key information, enabling it to handle any number of time signatures and key signatures in a piece, without being limited to specific notations such as C major and 4/4.
    \item The AutoHarmonizer predicts chords frame-by-frame, which enables the generation of flexible harmonic rhythms.
    \item The utilization of gamma sampling allows users to adjust the harmonic density of model-generated chord progressions.
\end{itemize}

\section*{Related Work}
\subsection*{Melody Harmonization}
Melody harmonization is a branch of algorithmic composition \cite{makris2016automatic}, which aims to generate a chord progression automatically for a given melody \cite{DBLP:conf/ismir/LimRL17,DBLP:journals/corr/abs-2001-02360,DBLP:conf/icassp/SunCLCW21}. Some of these studies have also focused on generating a four-part chorale to accompany a given melody \cite{DBLP:conf/ismir/LiangG0S17,DBLP:conf/icml/HadjeresPN17,DBLP:journals/corr/abs-1907-06637}. The present paper specifically addresses the former approach.

Tsushima et al. \cite{DBLP:conf/ismir/TsushimaNIY17} proposed a method for chord hierarchy representation based on a Probabilistic Context-Free Grammar (PCFG) of chords. They developed a metrical Markov model for controllable chord generation using this hierarchical representation. However, this approach relies on statistical learning, which tends to generate simpler and more basic chord sequences, resulting in fewer generated chords than bars.

Lim et al. \cite{DBLP:conf/ismir/LimRL17} designed a model based on a Bi-directional Long and Short-Term Memory (Bi-LSTM) network that can generate a chord from 24 triads for each bar. However, this model has limitations such as disregarding note order, rhythm, and octave information within bars. It generates results with overuse of common chords and inappropriate cadences.

\begin{figure*}[t]
    \centering
        \begin{minipage}{12cm} 
            \includegraphics[width=\textwidth]{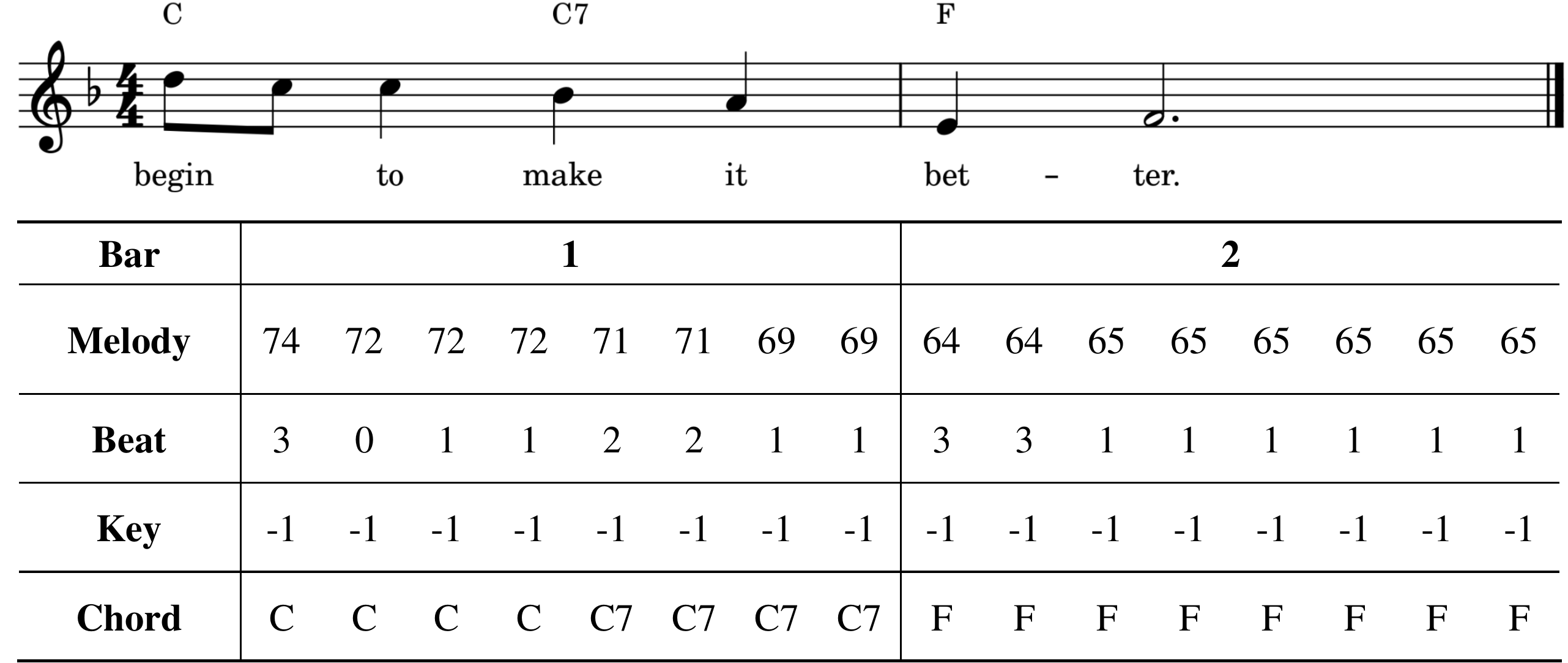}
        \end{minipage}
    \vspace{1em}
    \caption{A two-bar sample of a melody, beat, key, and chord representation (at a time resolution of eighth notes).
    \centering}
\end{figure*}

To address these limitations, Yeh et al. \cite{DBLP:journals/corr/abs-2001-02360} extended Lim's model, called MTHarmonizer, to predict a chord from 48 triads for each half-bar. They also included some extra information such as tonic and dominant to improve the model's performance.

In \cite{DBLP:conf/icassp/SunCLCW21}, Sun et al. applied orderless sampling and class weighting to the Bi-LSTM model. They expanded the types of chords to 96, and subjective experiments demonstrated that the generated results were comparable to those produced by human composers.

Chen et al. \cite{DBLP:conf/ismir/ChenLCW21} proposed SurpriseNet, a model based on a Conditional Variational Auto-Encoder (CVAE) and Bi-LSTM. This model enables user-controllable melody harmonization.

Yang et al. \cite{yang2019clstms} utilized two LSTM models. One model focused on the relationship between notes in the melody and their corresponding chords, while the other model focused on the rules of chord transfer.

Majidi et al. \cite{DBLP:journals/corr/abs-2102-07960} combined genetic algorithms with LSTMs to generate and optimize melodies and chords.

A recent deep learning approach \cite{9723052} by Rhyu et al. leverages a Transformer architecture, combined with a VAE framework, to generate structured chord sequences from melodies.

It should be noted that all of the models mentioned above, except for Tsushima et al. \cite{DBLP:conf/ismir/TsushimaNIY17}, cannot generate flexible harmonic rhythms.

\subsection*{Controllable Music Generation}
Controllable music generation systems refer to computer programs or algorithms that can generate music based on specific requirements set by the user. These systems rely on the representation of various properties of music, which may be subjective, such as emotion and style, or objective, such as tonality and beat. The generation of music by these systems can be customized to meet the specific needs of the user and can be tailored to a particular application.

Controllable music generation has been an active research area, and several models have been proposed to achieve this goal. Roberts et al. proposed a model based on recurrent Variational Auto-Encoders (VAEs) \cite{46809}. This model enables controllable generation through hierarchical decoders, allowing for control over various musical features such as harmony, melody, and rhythm.

Luo et al. proposed a model based on VAEs with Gaussian mixture latent distributions \cite{DBLP:conf/ismir/LuoAH19}. This model enables the learning of decoupled representations of timbres and pitches, facilitating the control of these two musical features separately. Zhang et al. proposed BUTTER \cite{zhang2020butter}, a representation learning model based on VAE, which can learn latent representations and cross-modal representations of music. This model allows for searching or generating corresponding music by inputting text, providing users with more convenient control over the generation process.

Chen et al. proposed Music SketchNet \cite{DBLP:conf/ismir/0021WBD20}, which uses VAE to decouple rhythmic and pitch contours, allowing for guided generation based on user-specified rhythms and pitches. Wang et al. proposed PianoTree VAE \cite{DBLP:conf/ismir/0008ZZJYXZ20}, which uses a Gated Recurrent Unit (GRU) to encode notes played simultaneously and map them to a latent space. This model achieves controllable generation of polyphonic music based on a tree structure.

Finally, Di et al. proposed the Controllable Music Transformer \cite{DBLP:conf/mm/DiJ0WZHLY21}, which achieves rhythmic consistency between video and background music. This model allows for the local control of rhythm while globally controlling the music genre and instruments.

It should be noted that these aforementioned models are designed for controllable music generation, and integrating their techniques into models without this capability might be challenging.

\section*{Methodology}
\subsection*{Data Representation}
Our research has yielded a novel data representation that includes crucial meta-information obtained from sheet music, namely the time signature and key signature. The data representation is illustrated in Fig. 2, and it involves encoding each lead sheet into four sequences, with equal lengths. This approach enables us to accurately capture the temporal and harmonic structure of the music. By including both time signature and key signature in the data representation, we can capture the rhythmic and harmonic patterns of the music comprehensively. Consequently, this enhances the accuracy and completeness of our music representation.

\begin{itemize}
    \item 
    \textbf{Melody Sequence}: we adopt a 128-dimensional one-hot vector encoding scheme to represent musical frames. Specifically, each frame is represented as a one-hot vector with 128 dimensions, where the time resolution is set at the level of sixteenth notes. The first dimension of the vector is reserved for representing rests, while the remaining 127 dimensions correspond to the unique pitches in the MIDI standard (excluding pitch 0).
    
    \item 
    \textbf{Beat Sequence}: a sequence of 4-dimensional vectors based on time signatures. It represents the beat strength of each frame in the melody sequence. Its values range from 0 to 3, corresponding to non-beat, weak, medium-weight, and strong beats. This sequence provides important information on the rhythmic structure of the melody.
    
    \item 
    \textbf{Key Sequence}: the encoding of keys is based on the number of sharps or flats associated with each key. Specifically, flats are assigned a numerical value ranging from -7 to -1, while sharps are assigned a numerical value ranging from 1 to 7. Keys with no sharps or flats are assigned a value of 0. In total, there are 15 possible types of key encoding based on this system.
    
    \item 
    \textbf{Chord Sequence}: there are 1,461 unique chord symbols were identified in our dataset. The first dimension is reserved for rests, leading to a one-hot vector representation of 1,462 dimensions for each chord.
\end{itemize}

\subsection*{Network Architecture}
In musical composition, a crucial aspect is the consideration of individual notes that comprise a given melody segment to effectively match it with a suitable chord progression. Generally, chords that incorporate notes already present in the melody, namely chord tones, are preferred. Nevertheless, there may be situations where several chords align with the current set of notes, necessitating the selection of the subsequent chord based on the upcoming notes of the melody. Thus, the selection of chords in musical composition involves a balance between the notes that are currently being played and those that are yet to come. This process is essential in creating a harmonious and coherent musical composition.

AutoHarmonizer is a model developed to capture music information bidirectionally. It employs a Bi-LSTM backbone network and an encoder-decoder architecture, as shown in Fig. 1. The model consists of two encoders, the melody encoder, and the meta-info encoder. The melody encoder takes a melody sequence as input, while the meta-info encoder takes a concatenated sequence of beat and key sequences. Both encoders have two stacked blocks, each comprising a Bi-LSTM layer with 256 units and a time-distributed layer with 128 units. The last hidden states of the encoders are concatenated, and the resulting vector is used as input to the decoder. The decoder is made up of three stacked layers, and its output layer has 1,462 units, which represent chord types. The chord symbols are generated autoregressively, frame-by-frame (sixteenth note) in the decoder. During training, the model used a dropout rate of 0.2, a batch size of 512, and early stopping with a patience of 20 epochs, as determined by empirical evaluation.

\subsection*{Controllable Harmonic Density}
In \cite{wu2022sampling}, Wu et al. proposed the use of gamma sampling to control the language models based on the assumption that certain attributes of the generated text have a close correlation with the number of occurrences of specific tokens. Gamma sampling provides a means to generate controllable text by scaling the probability of the token associated with the attribute during the generation process:

\begin{equation}
\begin{aligned}
    p_{\mathcal{A}_{out}}&=p_{\mathcal{A}_{in}}^{tan(\frac{\pi \Gamma}{2})}, \\
    p_{a_{out}}&=p_{a_{in}}\cdot \frac{p_{\mathcal{A}_{out}}}{p_{\mathcal{A}_{in}}},\quad \forall a\in \mathcal{A}, \\
    p_{n_{out}}&= p_{n_{in}} \cdot (1 + \frac{p_{\mathcal{A}_{in}}-p_{\mathcal{A}_{out}}}{p_{\backslash \mathcal{A}_{in}}}),\quad \forall n\notin \mathcal{A},\\
\end{aligned}
\end{equation}

\noindent
where $\Gamma\in[0,1]$ is the user-controllable control strength, $\mathcal{A}$ is the set of attribute-related tokens (${\backslash \mathcal{A}}$ is its complement), $p_{a_{in/out}}$ is the input/out probability of an attribute-related token $a$, and the same goes for every non-attribute-related token $n$. When $\Gamma=0.5$, there is no change in the probability distribution, while when $\Gamma<0.5$, the probabilities of the attribute-related tokens increase and vice versa.

AutoHarmonizer uses a strategy to achieve controllable harmonic density whereby the previously generated chord token $c_{t-1}$ is selected as the attribute-related token for generating the chord $c_{t}$ at time step $t$. Specifically, when the value of a parameter called $\Gamma$ exceeds 0.5, the model is more inclined to generate chords that differ from $c_{t-1}$, resulting in a greater frequency of chord switching and an increase in harmonic density. Conversely, when $\Gamma$ is less than 0.5, the likelihood of generating denser chord progressions is reduced. This approach enables AutoHarmonizer to produce musical pieces with controllable harmonic complexity, allowing for flexibility in the generation of diverse music styles.

Musical composition typically involves the frequent usage of essential chords, notably the tonic, dominant, and subdominant chords. While it is true that a model can loop between these chords, our model, with the introduction of high values of $\Gamma$, tends to diversify its chord choices, resulting in a more even distribution of chord types. As demonstrated in Fig. 4, a higher $\Gamma$ leads to a more balanced distribution of scale degrees in the generated chord progressions, indicating the inclusion of less common chords. Thus, by adjusting $\Gamma$, composers can influence the harmonic density and the distribution between essential and non-essential chords, offering them more control over the musical texture and composition.

\section*{Experiments}
\subsection*{Setups}

\subsubsection*{Dataset}
In our study, we utilized Wikifonia.org's lead sheet dataset, consisting of 6,675 compositions predominantly from western genres such as rock, pop, country, jazz, folk, R\&B, and children's songs. In order to improve the quality of the dataset, we removed lead sheets that lacked chord symbols or did not switch chords within 4 bars, ultimately resulting in a subset of 5,204 lead sheets. Subsequently, we divided the subset into a training set comprising 90\% of the data and a validation set containing the remaining 10\%.

\subsubsection*{Baselines}
In our research, we chose two previous melody harmonization systems as our baselines. The first is a traditional melody harmonization system proposed by Lim et al. \cite{DBLP:conf/ismir/LimRL17}, known as Chord Generation from Symbolic Melody (CGSM). This system is based on a Bi-LSTM architecture and was trained and validated using the Wikifonia dataset. The other two baseline models are STHarm and VTHarm \cite{9723052}. Both of these models were trained on the Chord Melody Dataset (CMD)\footnote{https://github.com/shiehn/chord-melody-dataset}. STHarm directly translates melodies into chords by mapping individual melody notes to chord progressions. On the other hand, VTHarm, featuring a key-aware variational Transformer architecture, not only generates chords from melodies but also captures the broader musical context and structure.

It is important to note that the strategy adopted by STHarm and VTHarm, which generates two chords per bar, has its limitations; they are only applicable to pieces in 4/4 time. Specifically, out of the 515 valid pieces in the validation set, only 311 are in 4/4 time. This implies that their comparison with other models might not necessarily be apple-to-apple.

We evaluated our system in various settings to determine the effectiveness of controllable harmonic density in melody harmonization. The proposed system, referred to as AH-$\Gamma$, consists of AutoHarmonizer set at different $\Gamma$ values ranging from 0.5 to 0.9. Through our analysis, we sought to establish a deeper understanding of the relationship between harmonic density and melody harmonization.

\subsection*{Metrics}
Our study evaluated the performance of AutoHarmonizer using a variety of metrics. These metrics included Accuracy (ACC), which measured the proportion of matching frames between the generated and true chord progressions. We also utilized six metrics proposed in a previous study \cite{DBLP:journals/corr/abs-2001-02360} that have become widely used in the literature \cite{DBLP:journals/access/RhyuCKL22,DBLP:conf/ismir/ChenLCW21,DBLP:conf/icassp/SunCLCW21} for assessing chord progression and melody/chord harmonicity. By utilizing these metrics, we were able to thoroughly evaluate the performance of AutoHarmonizer.

\begin{table*}[t]
    \centering
        \caption{Quantitative evaluations on the validation set (515 tunes). The values closest to the ground truth are bolded.
    \centering}
    \centering
    \centering
        \begin{minipage}{15cm} 
            \includegraphics[width=\textwidth]{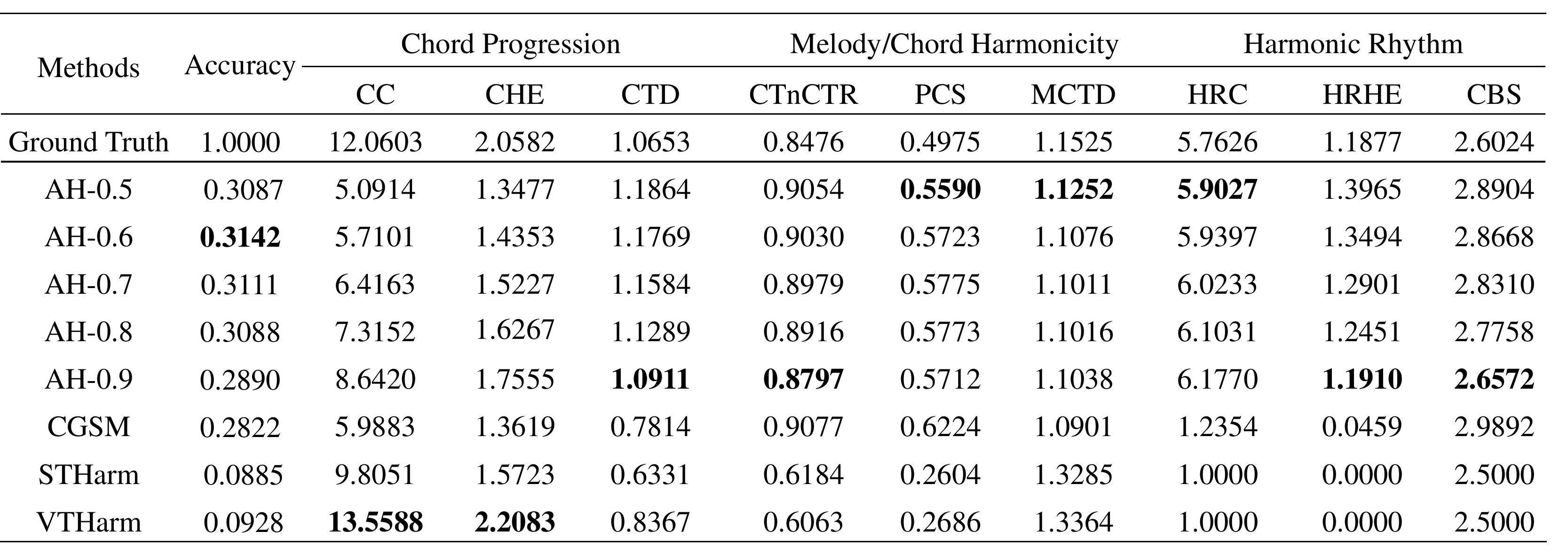}
        \end{minipage}
    \centering
\end{table*}

\begin{itemize}
\item 
    \textbf{Chord Coverage (CC)}: the number of chord types in a piece of music. This value serves as an indicator of the richness and variety of chord progressions in the music, with higher CC values indicating a greater number of distinct chord types being used.
    
    \item 
    \textbf{Chord Histogram Entropy (CHE)}: creates a histogram of chord occurrences based on a chord sequence:
    \begin{equation}
    \begin{aligned}
        CHE=-\sum_{k=1}^{CC}p_{k} \cdot {\rm log}p_{k},
    \end{aligned}
    \end{equation}
    where $p_{k}$ is the frequency of the $k$-th chord occurrence. The higher the value of CHE, the greater the uncertainty and the variety of chords.
    
    \item 
    \textbf{Chord Tonal Distance (CTD)}: the average value of the tonal distance \cite{harte2006detecting} computed between every pair of adjacent chords in a given chord sequence. It involves three steps: 1) the Pitch Class Profile (PCP) features of both chords are computed; 2) these features are then projected onto a six-dimensional tonal space; 3) the Euclidean distance between the two six-dimensional feature vectors is calculated, resulting in a tonal distance value. The lower the value of CTD, the smoother the chord progression.
    
    \item 
    \textbf{Chord Tone to non-Chord Tone Ratio (CTnCTR)}: calculates the ratio of the number of the chord tones ($n_{c}$) and proper non-chord tones ($n_{p}$), to the number of the non-chord tones ($n_{n}$):
    \begin{equation}
    \begin{aligned}
        CTnCTR=\frac{n_{c}+n_{p}}{n_{c}+n_{n}},
    \end{aligned}
    \end{equation}
    The concept of chord tones refers to melody notes whose pitch class belongs to the current chord, specifically, one of the three pitch classes that constitute a triad for the corresponding half bar. Melody notes that do not fall into this category are considered non-chord tones. Among the non-chord tones, a subset of notes that are two semitones away from the notes immediately following them is referred to as proper non-chord tones. CTnCTR equals one when there are no non-chord tones at all, or all non-chord tones are proper.
    
    \item 
    \textbf{Pitch Consonance Score (PCS)}: based on the musical interval between the pitch of the melody note and the chord notes, assuming that the pitch of the melody notes is always higher, which is the case in our system. Specifically, it assigns a score of 1 to consonant intervals, including unison, major/minor 3rd, perfect 5th, and major/minor 6th. A perfect 4th receives a score of 0, while other intervals are considered dissonant and receive a score of -1. To compute PCS for a pair of melody and chord sequences, these consonance scores are averaged across 16th-note windows, excluding rest periods.
    
    \item 
    \textbf{Melody-Chord Tonal Distance (MCTD)}: represents a melody note using a PCP feature vector, which is essentially a one-hot vector. Next, it compares the PCP of this vector against the PCP of a chord label in a 6-D tonal space \cite{harte2006detecting}. The resulting measure provides an estimate of the closeness between the melody note and the chord label. To obtain a comprehensive measure of tonal distance between a melody sequence and its corresponding chord labels, its calculates the average of the tonal distance between every melody note and the corresponding chord label across the melody sequence. It also weights each distance by the duration of the corresponding melody note.
    
\end{itemize}

In light of the fact that the previously mentioned metrics do not consider the measurement of harmonic rhythm, we developed an additional set of three metrics, as outlined below.

\begin{itemize}
    \item 
    \textbf{Harmonic Rhythm Coverage (HRC)}: similar to CC, but it is computed specifically for harmonic rhythm types. A higher HRC value indicates the use of a greater number of unique harmonic rhythm types.
    
    \item 
    \textbf{Harmonic Rhythm Histogram Entropy (HR-HE)}: same as CHE, but calculates the histogram of harmonic rhythm:
    \begin{equation}
    \begin{aligned}
        HRHE=-\sum_{u=1}^{HRC}p_{u} \cdot {\rm log}p_{u},
    \end{aligned}
    \end{equation}
    where $p_{u}$ is the frequency of the $u$-th harmonic rhythm. The HRHE value reflects the uncertainty and variety of harmonic rhythms in the music. A higher HRHE value indicates greater variation and uncertainty in the harmonic rhythms.
    
    \item 
    \textbf{Chord Beat Strength (CBS)}: chord placements can be assessed by their average beat strength, which is scored on a scale ranging from 0 (non-beat) to 3 (strong beat), based on the beat sequence. The beat sequence determines the recurring pulse of a musical composition and is used to evaluate the placement of chords in relation to the underlying beat. A smaller value of the CBS metric indicates that more chords are positioned on non-strong beats, while a higher CBS value indicates the opposite. Therefore, CBS provides a quantitative measure of the degree to which chords are aligned with the underlying beat.
\end{itemize}

\begin{figure*}[t]
    \centering
        \begin{minipage}{11cm} 
            \includegraphics[width=\textwidth]{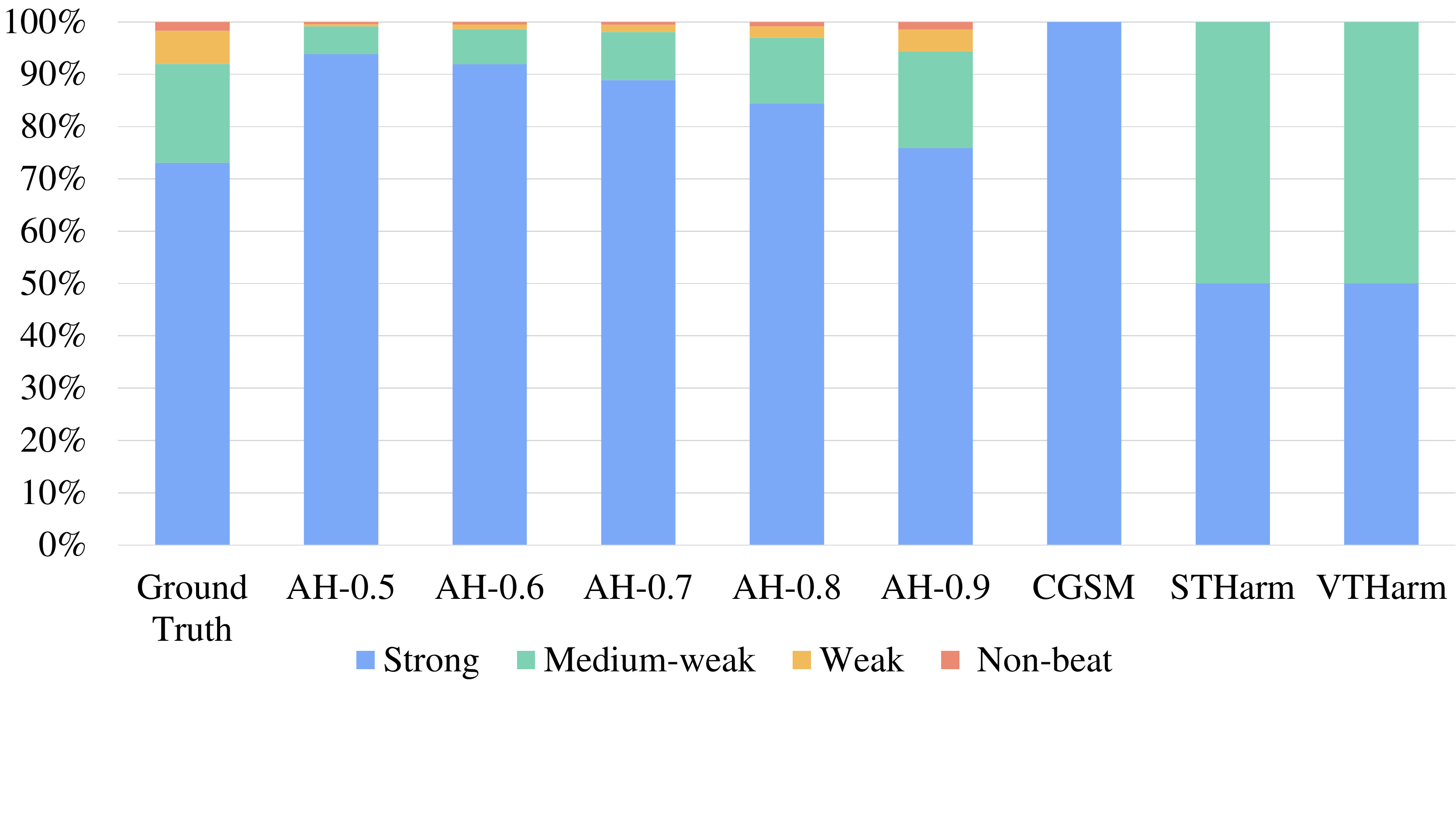}
        \end{minipage}
        \vspace{-3em}
    \caption{Distribution of chord onsets on beat strengths, showing the proportion of chord onsets occurring on strong, medium-weak, and weak beats, and non-beats.}
\end{figure*}

\begin{figure*}[t]
    \centering
        \begin{minipage}{14cm} 
            \includegraphics[width=\textwidth]{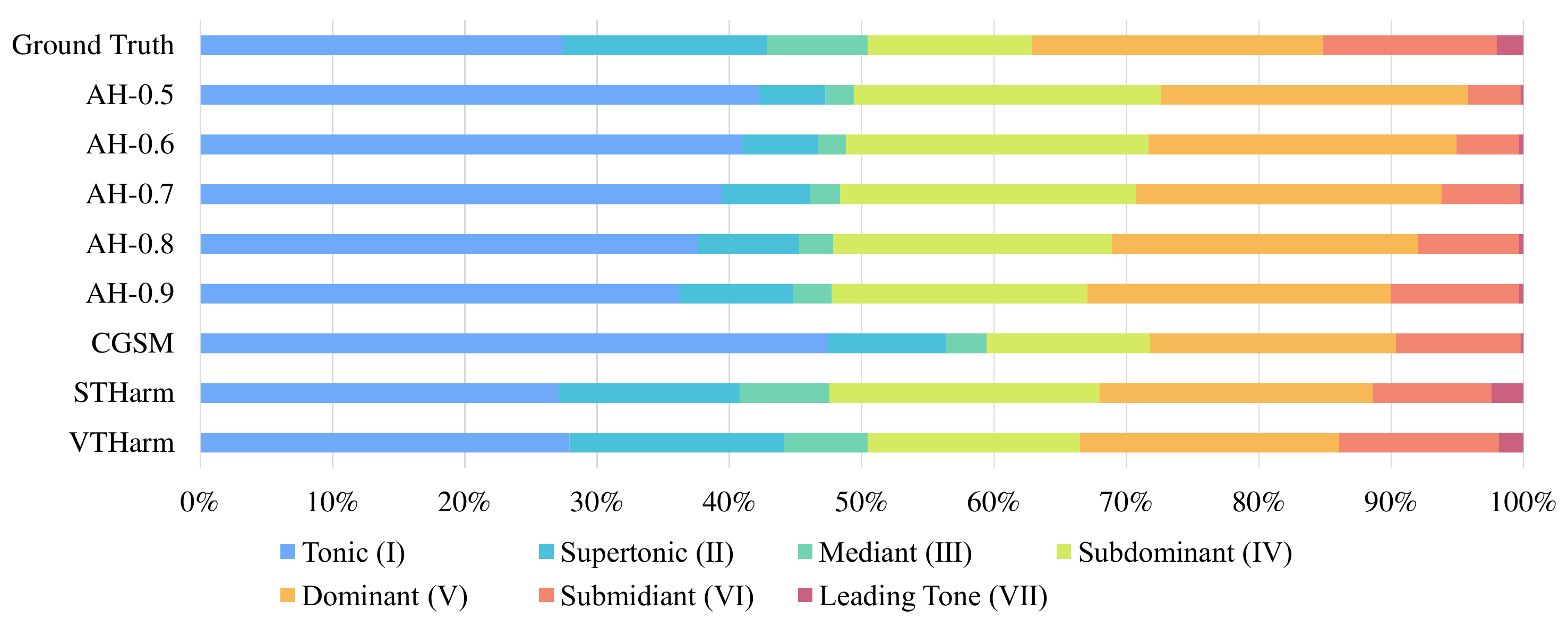}
        \end{minipage}
    \caption{Distribution of scale degrees in chord progressions generated by AutoHarmonizer and baselines, compared to chord progressions from ground truth.}
\end{figure*}

\subsection*{Quantitative Evaluations}
Tab. 1 presents the findings of this study, which can serve as a foundation for further analysis. It is crucial to note, however, that the metrics employed in this study are not meant to provide an all-encompassing evaluation of chord progression quality, as such assessments are inherently intricate and subjective \cite{DBLP:journals/corr/abs-2001-02360}. Therefore, although the results presented in Tab. 1 can be used for comparison purposes, they should not be viewed as definitive indicators of chord progression quality. It is imperative to acknowledge these limitations when interpreting the findings of this study.

\begin{figure*}[t]
    \centering
        \begin{minipage}{16.5cm} 
            \includegraphics[width=\textwidth]{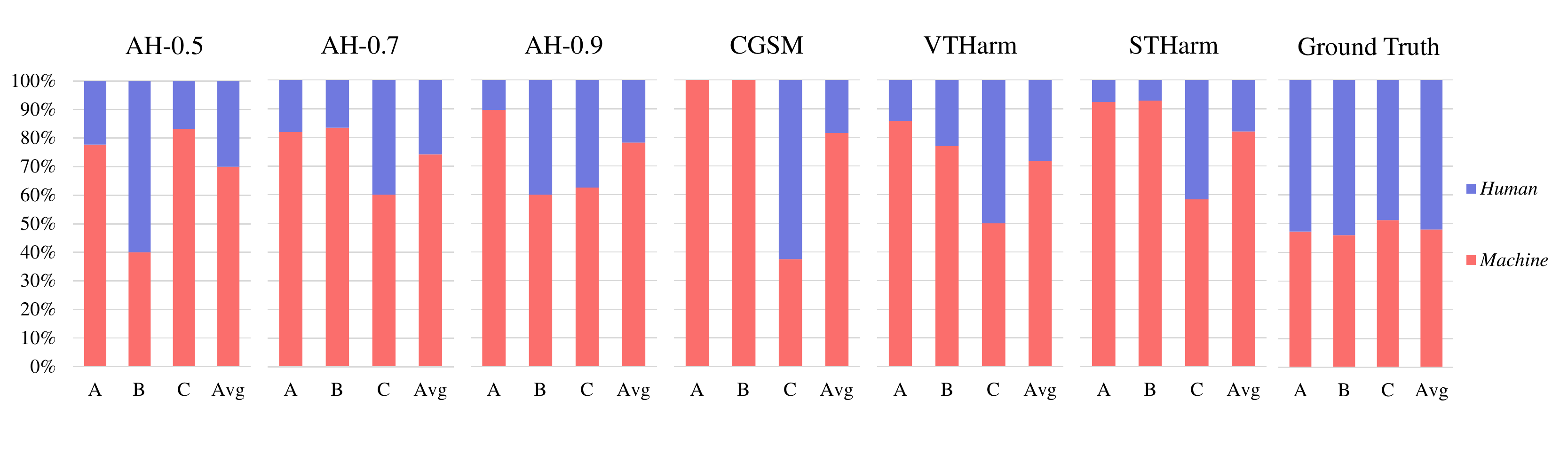}
        \end{minipage}
        \vspace{-2em}
        \caption{Results of the discrimination test for each model and expertise level. Group A: Music faculty and students; Group B: Non-music majors with harmony knowledge; Group C: Individuals without harmony knowledge but frequent music listeners. The term "vote" refers to the participants' tendency to categorize a given set of chords as either human-generated or machine-generated.}
    \centering
\end{figure*}

The results of ACC demonstrate that AutoHarmonizer consistently surpassed CGSM, VTHarm, and STHarm across all values of $\Gamma$. One might initially conclude that without incorporating less frequent chords, an over-reliance on prevalent triads could fail to effectively encapsulate the subtleties of human-composed musical compositions. However, it's crucial to note that the comparatively lower ACC of VTHarm and STHarm may stem from disparities in their training data compared to the other models. Furthermore, we found that an increase in $\Gamma$ was associated with a decline in accuracy. This observation indicates that the increase in the frequency of chord transitions leads to a greater deviation from actual chord progressions. While our metrics are single-faceted and may not capture the entirety of what makes a composition human-like, they do emphasize the potential benefit of including a variety of chord progressions in generation.

The chord progression metrics, CC, CHE, and CTD, serve as essential tools for evaluating both ground truth and model-generated chord progressions. Ground truth show low CTD, high CC, and CHE, indicating inherent smoothness and diversity. Among model-generated progressions, STHarm and VTHarm closely aligns with ground truth for CC and CHE. However, its considerably lower CTD is a result of frequently allocating identical chords within bars, leading to an `pseudo-smoothness'. Unlike VTHarm, the data representation of AH prevents consecutive identical chords, making AH-0.9's CTD closely resemble the ground truth. Meanwhile, CGSM offers the smoothest progressions with the lowest CTD but sacrifices diversity, evidenced by its low CC and CHE. This highlights the trade-off between progression smoothness and diversity.

The results of melody/chord harmonicity indicate that the actual musical compositions exhibit a greater prevalence of non-chord tones (the lowest CTnCTR), resulting in intervals that are more dissonant (the lowest PCS and highest MCTD) compared to those produced by the model. Notably, as the value of $\Gamma$ increased, the AutoHarmonizer utilized more non-chord tones while also forming more consonant intervals with the melody notes, resulting in decreased MCTD and increased PCS, suggesting a tendency towards consonance. In addition, both STHarm and VTHarm showed low CTnCTR and PCS values, with high MCTD, indicating a more pronounced deviation from the ground truth. These findings indicate that there may be significant differences in harmonic structures between human-composed and model-generated musical compositions (see Fig. 6 for examples of melody harmonization by various models).

The harmonic rhythm metrics demonstrate a clear contrast between chord progressions generated by the CGSM and those produced by AutoHarmonizer and ground truth. Specifically, chord progressions generated by the CGSM, STHarm, and VTHarm exhibit a fixed harmonic rhythm, which is a shared limitation among all melody harmonization neural network systems \cite{DBLP:journals/corr/abs-2001-02360,DBLP:conf/icassp/SunCLCW21,DBLP:conf/ismir/ChenLCW21}. The results suggest that an increase in the parameter $\Gamma$ leads to greater rhythmic diversity of chords, although the distribution is more concentrated, resulting in higher values of HRC and lower HRHE. Furthermore, an increase in $\Gamma$ leads to the placement of more chords on non-strong beats, as demonstrated by the lower values of CBS. These findings indicate that the parameter $\Gamma$ plays a critical role in generating varied and complex chord progressions with a non-fixed harmonic rhythm.

\begin{figure*}[t]
    \centering
        \begin{minipage}{15cm} 
            \includegraphics[width=\textwidth]{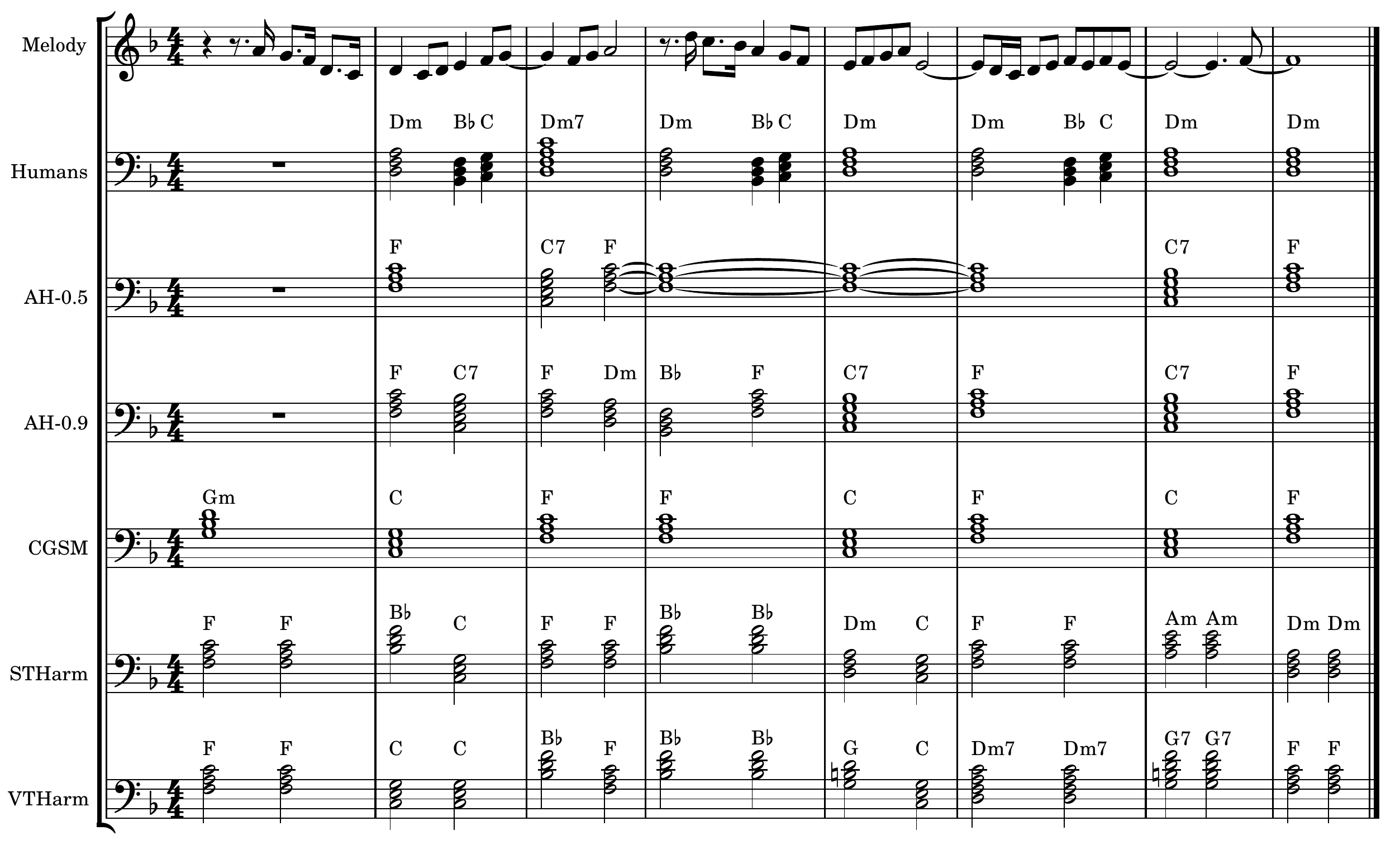}
        \end{minipage}
        \vspace{1em}
        \caption{Examples of melody harmonization. They are generated by various models from the same melody.
    \centering}
    \centering
    
\end{figure*}

Figure 3 depicts the distribution of chord onsets on beat strengths. It is evident from the figure that most chord onsets occur on strong beats, while a lesser number of chord onsets occur on medium-weak beats, and a small number of chord onsets occur on weak beats. Non-beats are seldom used for chord placement. Notably, CGSM restricts chord progression variety by exclusively positioning chords on strong beats. In a similar vein, both STHarm and VTHarm distribute their chords evenly between strong and medium-weak beats, indicating a potential monotony in their transitions. In contrast, by increasing the $\Gamma$ parameter in AutoHarmonizer, there is a shift towards placing chord onsets on non-strong beats. Among these, the AH-0.9 model comes closest to mirroring the ground truth's chord onset distribution.

The distribution of scale degrees in chord progressions generated by AutoHarmonizer and baselines is depicted in Fig. 4. The results highlight a consistent pattern across all models, with the majority of chords being tonic, followed by subdominant and dominant chords. However, it's essential to note that some models, such as CGSM, tend to overuse tonic chords compared to the ground truth. In contrast, STHarm and VTHarm exhibit a distribution remarkably close to the ground truth. Notably, STHarm's utilization of scale degrees is even more evenly spread than that of the ground truth, indicating a diverse array of generated chord progressions. Interestingly, when we increase the parameter $\Gamma$, there is a noticeable decrease in the usage of tonic chords, accompanied by a shift towards a more prominent presence of supertonic and submediant chords. This adjustment results in chord progressions that closely resemble the ground truth. By tuning the value of $\Gamma$, AutoHarmonizer can achieve a more balanced distribution across all chord degrees, thereby enhancing the diversity and richness of the generated chord progressions.

\subsection*{Discrimination Test}
This study recruited 83 participants with varying cultural and demographic backgrounds to engage in a discrimination task that aimed to differentiate between musical chords generated by humans and those created by machines.

The study categorized the subjects into three groups based on their level of music knowledge: A) Music faculty and students consisting of 33 subjects; B) Non-music majors with an understanding of harmony comprising 25 subjects; and C) Individuals with no knowledge of harmony but who listen to music often or occasionally, consisting of 25 subjects. The study selected a total of 20 tunes randomly from the validation set. Each tune had seven different versions, which included AH-0.5, AH-0.7, AH-0.9, CGSM, STHarm, VTHarm, and the ground truth.

It is important to note that the 20 tunes used for the listening test were randomly selected from the validation set. This choice was made because our dataset does not explicitly has a dedicated test set. We believe that this selection does not compromise the fairness of our experiments. Our primary focus is on comparing the performance of different models, rather than tuning the models.

Participants were presented with both melody-only versions and the ones with chords, and their task was to determine whether the chords were produced by humans or machines. The test consisted of questions that had an equal chance of containing chords generated by either humans (i.e., the ground truth) or machines. The primary objective of this study was to assess the capacity of the machine-generated chords to deceive human listeners into believing they were created by humans, thereby determining the model's overall trustworthiness. The study design was also intended to provide insight into the ability of individuals from different backgrounds to distinguish between human and machine-generated musical content.

The results of the discrimination test, depicted in Fig. 5, demonstrate that AutoHarmonizer surpassed CGSM and STHarm in all evaluated configurations. Professional subjects possessing musical backgrounds A and B consistently identified the chord progressions generated by CGSM as machine-generated and nearly all those by STHarm in the same manner. This observation aligns with the limitations of CGSM, which are characterized by a restricted vocabulary size (24 types of chords only) and a fixed harmonic rhythm. The fixed harmonic rhythm often leads to repetitive chord progressions, which, in turn, can be perceived as less convincing by listeners. Similarly, STHarm utilizes a constant harmonic rhythm, echoing these shortcomings. VTHarm preserved this harmonic rhythm setup, but incorporated a key-aware context encoder, enhancing its output. Conversely, chord progressions produced by AH-0.5 were viewed as most akin to human-composed ones, based on subject feedback. Intriguingly, half the subjects mistook the authentic human compositions as machine-generated, highlighting a deep-rooted skepticism towards human-composed chord progressions. Furthermore, we noted that, in contrast to professionals (i.e., subjects in group A and B), individuals in group C were more inclined to classify progressions with fixed harmonic rhythms (i.e., CGSM, VTHarm, and STHarm) as human-composed. We speculate that they might deem progressions with a constant harmonic rhythm as more familiar or acceptable, especially since many popular music pieces employ recurrent and predictable chord patterns.

\section*{Melody Harmonization Examples}

Figure 6 showcases several instances of melody harmonization, which aims to exhibit the effectiveness of diverse techniques utilized in harmonizing an identical 8-bar melody that features a 4/4 time signature and is based in the key of F.

The progression that is manually created by humans shows a distinct periodicity, wherein each cycle is comprised of four chords, namely \texttt{Dm-Bb-C-Dm7}, with \texttt{Dm} functioning as the tonic chord. The progression conforms to a particular tonal structure, emphasizing the significant role played by the tonic chord in shaping the overall musical structure. The periodicity of the chord progression, with its repetition of chord progressions, provides a sense of predictability and familiarity that contributes to its aesthetic appeal. Furthermore, the use of the tonic chord creates a sense of resolution and stability, which is a common characteristic of tonal music. Thus, the chord progression crafted by humans showcases the creative potential of human composers in shaping musical structures through the careful selection and arrangement of chords.

The chord progression produced by AH-0.5 exhibited in this output is characterized by a significantly sparse harmonic rhythm, which is composed of only five chords in total. This chord progression follows a consistent pattern of \texttt{F-C7-F} in each cycle. The tonic chord is represented by \texttt{F}, indicating that the key of this chord progression is in F-major, which is the relative key of the ground truth. It is worth noting that for this melody, both D-minor and F-major chord progressions are considered acceptable according to principles in music theory.

In music theory, chord progressions are often evaluated based on their harmonic function and tonal relationships within a key. While the ground truth chord progression is in D minor, which is the relative minor key of F major, the use of the F major chord progression generated by AH-0.5 is considered acceptable because it represents the relative major key of D minor. The relative major and minor keys share many harmonically related chords, making both progressions suitable for harmonizing the same melody.

It can be observed that increasing the parameter $\Gamma$ to 0.9 resulted in an increase in both the density and variety of chord progressions generated by AutoHarmonizer. The generated chord progressions demonstrated two distinct types of cycles: \texttt{F-C7-F}, which consists of a tonic chord followed by a dominant chord and was also observed in the one generated by AH-0.5, and \texttt{F-Dm-Bb-F}, which includes a subdominant and supertonic chord, adding further movement to the progression. The increase in density and variety of chord progressions suggests that adjusting the $\Gamma$ parameter has a significant impact on the output of AutoHarmonizer and can be a valuable tool for musicians and composers seeking to explore different chord progressions.

The chord progression generated by CGSM differs from the others, as it employs only triads, given that this system's vocabulary does not include other chord types. Notably, the first bar is a pick-up bar (the bar before the first full bar), yet CGSM inserts a \texttt{Gm} chord in this position. This choice is unconventional since \texttt{Gm} does not correspond to a tonic, dominant, or subdominant chord. This is followed by \texttt{C-F} repeatedly, suggesting a simple alternation between dominant and tonic. The absence of other chord variations limits the progression's expressiveness and dynamic movement.

On the other hand, chord progressions generated by STHarm display a more diverse set of chord choices, extending beyond basic triads. The presence of \texttt{F}, \texttt{Bb}, and \texttt{C} chords indicate a familiarity with the key of F major and its respective harmonic nuances. Furthermore, the progression’s inclusion of the \texttt{Am} chord adds an interesting color to the harmony, introducing a bit more diversity to the harmonic landscape. However, while its choices are more varied than those of CGSM, STHarm's progression sometimes lacks the structural cohesion and clarity seen in the human-composed and AH-generated harmonizations.

VTHarm, compared to STHarm, presents more transitions, specifically with the shift from \texttt{F} to \texttt{C}, then introducing \texttt{Bb} and then progressing to \texttt{Dm7}, eventually transitioning through \texttt{G7} before resolving to \texttt{F}. These transitions indicate a more complex harmonic understanding than STHarm and certainly CGSM. However, its outputs occasionally diverge from traditional harmonic conventions, resulting in progressions that might be viewed as less cohesive by those familiar with music theory.

By showcasing these examples, we emphasize the nuances of each method, providing a comprehensive understanding of the strengths and weaknesses of different harmonization techniques. Additionally, we suggest that those interested can access and evaluate more generations of machine-generated chord progressions on GitHub\footnote{\url{https://sander-wood.github.io/autoharmonizer/test}}.

\section*{Conclusions}
AutoHarmonizer is a novel system for melody harmonization that aims to provide greater control over harmonic density and the flexibility of harmonic rhythm. The system has the ability to generate chord progressions with varying harmonic densities, thereby enabling users to create desirable chord progressions.

In order to evaluate the performance of AutoHarmonizer, a series of experiments were conducted. The results of these experiments demonstrate the effectiveness of the system in producing chord progressions with varying harmonic rhythms. The system also enables users to control the harmonic rhythm, thereby providing greater flexibility in creating unique chord progressions.

Despite the promising results, the evaluation of AutoHarmonizer using both quantitative metrics and a discrimination test suggests that there is still room for improvement in the quality of chord progressions generated by the system. Further work is required to address these limitations and enhance the overall performance of the system.

\begin{backmatter}
\section*{Abbreviations}%% if any
\begin{tabbing}
\hspace{3cm}\=\kill % 设置制表位宽度为 3cm
ACC \> Accuracy\\
AH \> AutoHarmonizer\\
CBS \> Chord Beat Strength\\
CC \> Chord Coverage\\
CGSM \> Chord Generation from Symbolic Melody\\
CHE \> Chord Histogram Entropy\\
CTD \> Chord Tonal Distance\\
CTnCTR \> Chord Tone to non-Chord Tone Ratio\\
GRU \> Gated Recurrent Unit\\
HRC \> Harmonic Rhythm Coverage\\
HRHE \> Harmonic Rhythm Histogram Entropy\\
LSTM \> Long and Short-Term Memory\\
MCTD \> Melody-Chord Tonal Distance\\
PCFG \> Probabilistic Context-Free Grammar\\
PCP \> Pitch Class Profile\\
PCS \> Pitch Consonance Score\\
VAE \> Variational Auto-Encoder
\end{tabbing}

\section*{Authors' contributions}
Shangda Wu initiated the research and wrote the paper, while Yue Yang and Zhaowen Wang contributed by investigating related work, conducting experiments, and writing corresponding parts. Xiaobing Li and Maosong Sun provided guidance and supervision throughout the project. All authors have reviewed and approved the final manuscript.

% \section*{Funding}%% if any
% This research work has been made possible by various funding sources, including the High-grade, Precision and Advanced Discipline Construction Project of Beijing Universities, the Major Projects of National Social Science Fund of China (Grant No. 21ZD19), and the Nation Culture and Tourism Technological Innovation Engineering Project.

\section*{Availability of data and materials}%% if any
The dataset, source code, model weights, and model-generated samples used in this project are publicly available at \url{https://github.com/sander-wood/autoharmonizer}.

%%  After submission of the .TEX file,                     %%
%%  you will be prompted to submit your .BBL file.         %%

% if your bibliography is in bibtex format, use those commands:
\bibliographystyle{bmc-mathphys} % Style BST file (bmc-mathphys, vancouver, spbasic).
\bibliography{bmc_article}      % Bibliography file (usually '*.bib' )


\begin{thebibliography}{30}
\ifx \bisbn   \undefined \def \bisbn  #1{ISBN #1}\fi
\ifx \binits  \undefined \def \binits#1{#1}\fi
\ifx \bauthor  \undefined \def \bauthor#1{#1}\fi
\ifx \batitle  \undefined \def \batitle#1{#1}\fi
\ifx \bjtitle  \undefined \def \bjtitle#1{#1}\fi
\ifx \bvolume  \undefined \def \bvolume#1{\textbf{#1}}\fi
\ifx \byear  \undefined \def \byear#1{#1}\fi
\ifx \bissue  \undefined \def \bissue#1{#1}\fi
\ifx \bfpage  \undefined \def \bfpage#1{#1}\fi
\ifx \blpage  \undefined \def \blpage #1{#1}\fi
\ifx \burl  \undefined \def \burl#1{\textsf{#1}}\fi
\ifx \doiurl  \undefined \def \doiurl#1{\textsf{#1}}\fi
\ifx \betal  \undefined \def \betal{\textit{et al.}}\fi
\ifx \binstitute  \undefined \def \binstitute#1{#1}\fi
\ifx \binstitutionaled  \undefined \def \binstitutionaled#1{#1}\fi
\ifx \bctitle  \undefined \def \bctitle#1{#1}\fi
\ifx \beditor  \undefined \def \beditor#1{#1}\fi
\ifx \bpublisher  \undefined \def \bpublisher#1{#1}\fi
\ifx \bbtitle  \undefined \def \bbtitle#1{#1}\fi
\ifx \bedition  \undefined \def \bedition#1{#1}\fi
\ifx \bseriesno  \undefined \def \bseriesno#1{#1}\fi
\ifx \blocation  \undefined \def \blocation#1{#1}\fi
\ifx \bsertitle  \undefined \def \bsertitle#1{#1}\fi
\ifx \bsnm \undefined \def \bsnm#1{#1}\fi
\ifx \bsuffix \undefined \def \bsuffix#1{#1}\fi
\ifx \bparticle \undefined \def \bparticle#1{#1}\fi
\ifx \barticle \undefined \def \barticle#1{#1}\fi
\ifx \bconfdate \undefined \def \bconfdate #1{#1}\fi
\ifx \botherref \undefined \def \botherref #1{#1}\fi
\ifx \url \undefined \def \url#1{\textsf{#1}}\fi
\ifx \bchapter \undefined \def \bchapter#1{#1}\fi
\ifx \bbook \undefined \def \bbook#1{#1}\fi
\ifx \bcomment \undefined \def \bcomment#1{#1}\fi
\ifx \oauthor \undefined \def \oauthor#1{#1}\fi
\ifx \citeauthoryear \undefined \def \citeauthoryear#1{#1}\fi
\ifx \endbibitem  \undefined \def \endbibitem {}\fi
\ifx \bconflocation  \undefined \def \bconflocation#1{#1}\fi
\ifx \arxivurl  \undefined \def \arxivurl#1{\textsf{#1}}\fi
\csname PreBibitemsHook\endcsname

%%% 1
\bibitem{DBLP:conf/mir/LiuZMHCZX21}
\begin{bchapter}
\bauthor{\bsnm{Liu}, \binits{A.}},
\bauthor{\bsnm{Zhang}, \binits{L.}},
\bauthor{\bsnm{Mei}, \binits{Y.}},
\bauthor{\bsnm{Han}, \binits{B.}},
\bauthor{\bsnm{Cai}, \binits{Z.}},
\bauthor{\bsnm{Zhu}, \binits{Z.}},
\bauthor{\bsnm{Xiao}, \binits{J.}}:
\bctitle{Residual recurrent {CRNN} for end-to-end optical music recognition on
  monophonic scores}.
In: \beditor{\bsnm{Liu}, \binits{B.}},
\beditor{\bsnm{Fu}, \binits{J.}},
\beditor{\bsnm{Chen}, \binits{S.}},
\beditor{\bsnm{Jin}, \binits{Q.}},
\beditor{\bsnm{Hauptmann}, \binits{A.G.}},
\beditor{\bsnm{Rui}, \binits{Y.}} (eds.)
\bbtitle{MMPT@ICMR2021: Proceedings of the 2021 Workshop on Multi-Modal
  Pre-Training for Multimedia Understanding, Taipei, Taiwan, August 21, 2021},
pp. \bfpage{23}--\blpage{27}.
\bpublisher{{ACM}}, \blocation{???}
(\byear{2021}).
doi:\doiurl{10.1145/3463945.3469056}.
\burl{https://doi.org/10.1145/3463945.3469056}
\end{bchapter}
\endbibitem

%%% 2
\bibitem{DBLP:conf/ismir/Calvo-ZaragozaR18}
\begin{bchapter}
\bauthor{\bsnm{Calvo{-}Zaragoza}, \binits{J.}},
\bauthor{\bsnm{Rizo}, \binits{D.}}:
\bctitle{Camera-primus: Neural end-to-end optical music recognition on
  realistic monophonic scores}.
In: \beditor{\bsnm{G{\'{o}}mez}, \binits{E.}},
\beditor{\bsnm{Hu}, \binits{X.}},
\beditor{\bsnm{Humphrey}, \binits{E.}},
\beditor{\bsnm{Benetos}, \binits{E.}} (eds.)
\bbtitle{Proceedings of the 19th International Society for Music Information
  Retrieval Conference, {ISMIR} 2018, Paris, France, September 23-27, 2018},
pp. \bfpage{248}--\blpage{255}
(\byear{2018}).
\burl{http://ismir2018.ircam.fr/doc/pdfs/33\_Paper.pdf}
\end{bchapter}
\endbibitem

%%% 3
\bibitem{DBLP:conf/interspeech/GhosalK18}
\begin{bchapter}
\bauthor{\bsnm{Ghosal}, \binits{D.}},
\bauthor{\bsnm{Kolekar}, \binits{M.H.}}:
\bctitle{Music genre recognition using deep neural networks and transfer
  learning}.
In: \beditor{\bsnm{Yegnanarayana}, \binits{B.}} (ed.)
\bbtitle{Interspeech 2018, 19th Annual Conference of the International Speech
  Communication Association, Hyderabad, India, 2-6 September 2018},
pp. \bfpage{2087}--\blpage{2091}.
\bpublisher{{ISCA}}, \blocation{???}
(\byear{2018}).
doi:\doiurl{10.21437/Interspeech.2018-2045}.
\burl{https://doi.org/10.21437/Interspeech.2018-2045}
\end{bchapter}
\endbibitem

%%% 4
\bibitem{DBLP:conf/evoW/DervakosKS21}
\begin{bchapter}
\bauthor{\bsnm{Dervakos}, \binits{E.}},
\bauthor{\bsnm{Kotsani}, \binits{N.}},
\bauthor{\bsnm{Stamou}, \binits{G.}}:
\bctitle{Genre recognition from symbolic music with cnns}.
In: \beditor{\bsnm{Romero}, \binits{J.}},
\beditor{\bsnm{Martins}, \binits{T.}},
\beditor{\bsnm{Rodr{\'{\i}}guez{-}Fern{\'{a}}ndez}, \binits{N.}} (eds.)
\bbtitle{Artificial Intelligence in Music, Sound, Art and Design - 10th
  International Conference, EvoMUSART 2021, Held as Part of EvoStar 2021,
  Virtual Event, April 7-9, 2021, Proceedings}.
\bsertitle{Lecture Notes in Computer Science},
vol. \bseriesno{12693},
pp. \bfpage{98}--\blpage{114}.
\bpublisher{Springer}, \blocation{???}
(\byear{2021}).
doi:\doiurl{10.1007/978-3-030-72914-1\_7}.
\burl{https://doi.org/10.1007/978-3-030-72914-1\_7}
\end{bchapter}
\endbibitem

%%% 5
\bibitem{DBLP:journals/corr/abs-1709-01620}
\begin{botherref}
\oauthor{\bsnm{Briot}, \binits{J.}},
\oauthor{\bsnm{Hadjeres}, \binits{G.}},
\oauthor{\bsnm{Pachet}, \binits{F.}}:
Deep learning techniques for music generation - {A} survey.
CoRR
\textbf{abs/1709.01620}
(2017)
\end{botherref}
\endbibitem

%%% 6
\bibitem{DBLP:conf/pimrc/CasiniMR18}
\begin{bchapter}
\bauthor{\bsnm{Casini}, \binits{L.}},
\bauthor{\bsnm{Marfia}, \binits{G.}},
\bauthor{\bsnm{Roccetti}, \binits{M.}}:
\bctitle{Some reflections on the potential and limitations of deep learning for
  automated music generation}.
In: \bbtitle{29th {IEEE} Annual International Symposium on Personal, Indoor and
  Mobile Radio Communications},
pp. \bfpage{27}--\blpage{31}.
\bpublisher{{IEEE}}, \blocation{???}
(\byear{2018})
\end{bchapter}
\endbibitem

%%% 7
\bibitem{DBLP:journals/csur/HerremansCC17}
\begin{barticle}
\bauthor{\bsnm{Herremans}, \binits{D.}},
\bauthor{\bsnm{Chuan}, \binits{C.}},
\bauthor{\bsnm{Chew}, \binits{E.}}:
\batitle{A functional taxonomy of music generation systems}.
\bjtitle{{ACM} Comput. Surv.}
\bvolume{50}(\bissue{5}),
\bfpage{69}--\blpage{16930}
(\byear{2017})
\end{barticle}
\endbibitem

%%% 8
\bibitem{makris2016automatic}
\begin{botherref}
\oauthor{\bsnm{Makris}, \binits{D.}},
\oauthor{\bsnm{Kayrdis}, \binits{I.}},
\oauthor{\bsnm{Sioutas}, \binits{S.}}:
Automatic melodic harmonization: An overview, challenges and future directions.
Trends in music information seeking, behavior, and retrieval for creativity,
146--165
(2016)
\end{botherref}
\endbibitem

%%% 9
\bibitem{DBLP:conf/icmcs/SunWY22}
\begin{bchapter}
\bauthor{\bsnm{Sun}, \binits{W.}},
\bauthor{\bsnm{Wu}, \binits{J.}},
\bauthor{\bsnm{Yuan}, \binits{S.}}:
\bctitle{Melodic skeleton: {A} musical feature for automatic melody
  harmonization}.
In: \bbtitle{{IEEE} International Conference on Multimedia and Expo Workshops,
  {ICME} Workshops 2022, Taipei, Taiwan, July 18-22, 2022},
pp. \bfpage{1}--\blpage{6}.
\bpublisher{{IEEE}}, \blocation{???}
(\byear{2022}).
doi:\doiurl{10.1109/ICMEW56448.2022.9859421}
\end{bchapter}
\endbibitem

%%% 10
\bibitem{DBLP:conf/ismir/TsushimaNIY17}
\begin{bchapter}
\bauthor{\bsnm{Tsushima}, \binits{H.}},
\bauthor{\bsnm{Nakamura}, \binits{E.}},
\bauthor{\bsnm{Itoyama}, \binits{K.}},
\bauthor{\bsnm{Yoshii}, \binits{K.}}:
\bctitle{Function- and rhythm-aware melody harmonization based on
  tree-structured parsing and split-merge sampling of chord sequences}.
In: \bbtitle{Proceedings of the 18th International Society for Music
  Information Retrieval Conference, {ISMIR} 2017, Suzhou, China, October 23-27,
  2017},
pp. \bfpage{502}--\blpage{508}
(\byear{2017})
\end{bchapter}
\endbibitem

%%% 11
\bibitem{DBLP:conf/ictai/BrunnerWWW17}
\begin{bchapter}
\bauthor{\bsnm{Brunner}, \binits{G.}},
\bauthor{\bsnm{Wang}, \binits{Y.}},
\bauthor{\bsnm{Wattenhofer}, \binits{R.}},
\bauthor{\bsnm{Wiesendanger}, \binits{J.}}:
\bctitle{Jambot: Music theory aware chord based generation of polyphonic music
  with lstms}.
In: \bbtitle{29th {IEEE} International Conference on Tools with Artificial
  Intelligence, {ICTAI} 2017, Boston, MA, USA, November 6-8, 2017},
pp. \bfpage{519}--\blpage{526}.
\bpublisher{{IEEE} Computer Society}, \blocation{???}
(\byear{2017}).
doi:\doiurl{10.1109/ICTAI.2017.00085}
\end{bchapter}
\endbibitem

%%% 12
\bibitem{DBLP:conf/icassp/SunCLCW21}
\begin{bchapter}
\bauthor{\bsnm{Sun}, \binits{C.}},
\bauthor{\bsnm{Chen}, \binits{Y.}},
\bauthor{\bsnm{Lee}, \binits{H.}},
\bauthor{\bsnm{Chen}, \binits{Y.}},
\bauthor{\bsnm{Wang}, \binits{H.}}:
\bctitle{Melody harmonization using orderless nade, chord balancing, and
  blocked gibbs sampling}.
In: \bbtitle{{IEEE} International Conference on Acoustics, Speech and Signal
  Processing, {ICASSP} 2021, Toronto, ON, Canada, June 6-11, 2021},
pp. \bfpage{4145}--\blpage{4149}.
\bpublisher{{IEEE}}, \blocation{???}
(\byear{2021}).
doi:\doiurl{10.1109/ICASSP39728.2021.9414281}
\end{bchapter}
\endbibitem

%%% 13
\bibitem{DBLP:journals/corr/abs-2001-02360}
\begin{botherref}
\oauthor{\bsnm{Yeh}, \binits{Y.}},
\oauthor{\bsnm{Hsiao}, \binits{W.}},
\oauthor{\bsnm{Fukayama}, \binits{S.}},
\oauthor{\bsnm{Kitahara}, \binits{T.}},
\oauthor{\bsnm{Genchel}, \binits{B.}},
\oauthor{\bsnm{Liu}, \binits{H.}},
\oauthor{\bsnm{Dong}, \binits{H.}},
\oauthor{\bsnm{Chen}, \binits{Y.}},
\oauthor{\bsnm{Leong}, \binits{T.}},
\oauthor{\bsnm{Yang}, \binits{Y.}}:
Automatic melody harmonization with triad chords: {A} comparative study.
CoRR
\textbf{abs/2001.02360}
(2020).
\arxivurl{2001.02360}
\end{botherref}
\endbibitem

%%% 14
\bibitem{DBLP:conf/ismir/ChenLCW21}
\begin{bchapter}
\bauthor{\bsnm{Chen}, \binits{Y.}},
\bauthor{\bsnm{Lee}, \binits{H.}},
\bauthor{\bsnm{Chen}, \binits{Y.}},
\bauthor{\bsnm{Wang}, \binits{H.}}:
\bctitle{Surprisenet: Melody harmonization conditioning on user-controlled
  surprise contours}.
In: \bbtitle{Proceedings of the 22nd International Society for Music
  Information Retrieval Conference, {ISMIR} 2021, Online, November 7-12, 2021},
pp. \bfpage{105}--\blpage{112}
(\byear{2021})
\end{bchapter}
\endbibitem

%%% 15
\bibitem{wu2022sampling}
\begin{botherref}
\oauthor{\bsnm{Wu}, \binits{S.}},
\oauthor{\bsnm{Sun}, \binits{M.}}:
Efficient and training-free control of language generation.
CoRR
\textbf{abs/2205.06036}
(2022)
\end{botherref}
\endbibitem

%%% 16
\bibitem{DBLP:conf/ismir/LimRL17}
\begin{bchapter}
\bauthor{\bsnm{Lim}, \binits{H.}},
\bauthor{\bsnm{Rhyu}, \binits{S.}},
\bauthor{\bsnm{Lee}, \binits{K.}}:
\bctitle{Chord generation from symbolic melody using {BLSTM} networks}.
In: \bbtitle{Proceedings of the 18th International Society for Music
  Information Retrieval Conference, {ISMIR} 2017, Suzhou, China, October 23-27,
  2017},
pp. \bfpage{621}--\blpage{627}
(\byear{2017})
\end{bchapter}
\endbibitem

%%% 17
\bibitem{DBLP:conf/ismir/LiangG0S17}
\begin{bchapter}
\bauthor{\bsnm{Liang}, \binits{F.T.}},
\bauthor{\bsnm{Gotham}, \binits{M.}},
\bauthor{\bsnm{Johnson}, \binits{M.}},
\bauthor{\bsnm{Shotton}, \binits{J.}}:
\bctitle{Automatic stylistic composition of bach chorales with deep {LSTM}}.
In: \bbtitle{Proceedings of the 18th International Society for Music
  Information Retrieval Conference},
pp. \bfpage{449}--\blpage{456}
(\byear{2017})
\end{bchapter}
\endbibitem

%%% 18
\bibitem{DBLP:conf/icml/HadjeresPN17}
\begin{bchapter}
\bauthor{\bsnm{Hadjeres}, \binits{G.}},
\bauthor{\bsnm{Pachet}, \binits{F.}},
\bauthor{\bsnm{Nielsen}, \binits{F.}}:
\bctitle{Deepbach: a steerable model for bach chorales generation}.
In: \bbtitle{Proceedings of the 34th International Conference on Machine
  Learning},
vol. \bseriesno{70},
pp. \bfpage{1362}--\blpage{1371}.
\bpublisher{{PMLR}}, \blocation{???}
(\byear{2017})
\end{bchapter}
\endbibitem

%%% 19
\bibitem{DBLP:journals/corr/abs-1907-06637}
\begin{botherref}
\oauthor{\bsnm{Huang}, \binits{C.A.}},
\oauthor{\bsnm{Hawthorne}, \binits{C.}},
\oauthor{\bsnm{Roberts}, \binits{A.}},
\oauthor{\bsnm{Dinculescu}, \binits{M.}},
\oauthor{\bsnm{Wexler}, \binits{J.}},
\oauthor{\bsnm{Hong}, \binits{L.}},
\oauthor{\bsnm{Howcroft}, \binits{J.}}:
The bach doodle: Approachable music composition with machine learning at scale.
CoRR
\textbf{abs/1907.06637}
(2019).
\arxivurl{1907.06637}
\end{botherref}
\endbibitem

%%% 20
\bibitem{yang2019clstms}
\begin{bchapter}
\bauthor{\bsnm{Yang}, \binits{W.}},
\bauthor{\bsnm{Sun}, \binits{P.}},
\bauthor{\bsnm{Zhang}, \binits{Y.}},
\bauthor{\bsnm{Zhang}, \binits{Y.}}:
\bctitle{Clstms: A combination of two lstm models to generate chords
  accompaniment for symbolic melody}.
In: \bbtitle{2019 International Conference on High Performance Big Data and
  Intelligent Systems (HPBD\&IS)},
pp. \bfpage{176}--\blpage{180}
(\byear{2019}).
\bcomment{IEEE}
\end{bchapter}
\endbibitem

%%% 21
\bibitem{DBLP:journals/corr/abs-2102-07960}
\begin{botherref}
\oauthor{\bsnm{Majidi}, \binits{M.}},
\oauthor{\bsnm{Toroghi}, \binits{R.M.}}:
Music harmony generation, through deep learning and using a multi-objective
  evolutionary algorithm.
CoRR
\textbf{abs/2102.07960}
(2021).
\arxivurl{2102.07960}
\end{botherref}
\endbibitem

%%% 22
\bibitem{9723052}
\begin{barticle}
\bauthor{\bsnm{Rhyu}, \binits{S.}},
\bauthor{\bsnm{Choi}, \binits{H.}},
\bauthor{\bsnm{Kim}, \binits{S.}},
\bauthor{\bsnm{Lee}, \binits{K.}}:
\batitle{Translating melody to chord: Structured and flexible harmonization of
  melody with transformer}.
\bjtitle{IEEE Access}
\bvolume{10},
\bfpage{28261}--\blpage{28273}
(\byear{2022}).
doi:\doiurl{10.1109/ACCESS.2022.3155467}
\end{barticle}
\endbibitem

%%% 23
\bibitem{46809}
\begin{bbook}
\beditor{\bsnm{Roberts}, \binits{A.}},
\beditor{\bsnm{Engel}, \binits{J.}},
\beditor{\bsnm{Eck}, \binits{D.}} (eds.):
\bbtitle{Hierarchical Variational Autoencoders for Music}
(\byear{2017})
\end{bbook}
\endbibitem

%%% 24
\bibitem{DBLP:conf/ismir/LuoAH19}
\begin{bchapter}
\bauthor{\bsnm{Luo}, \binits{Y.}},
\bauthor{\bsnm{Agres}, \binits{K.}},
\bauthor{\bsnm{Herremans}, \binits{D.}}:
\bctitle{Learning disentangled representations of timbre and pitch for musical
  instrument sounds using gaussian mixture variational autoencoders}.
In: \beditor{\bsnm{Flexer}, \binits{A.}},
\beditor{\bsnm{Peeters}, \binits{G.}},
\beditor{\bsnm{Urbano}, \binits{J.}},
\beditor{\bsnm{Volk}, \binits{A.}} (eds.)
\bbtitle{Proceedings of the 20th International Society for Music Information
  Retrieval Conference, {ISMIR} 2019, Delft, The Netherlands, November 4-8,
  2019},
pp. \bfpage{746}--\blpage{753}
(\byear{2019}).
\burl{http://archives.ismir.net/ismir2019/paper/000091.pdf}
\end{bchapter}
\endbibitem

%%% 25
\bibitem{zhang2020butter}
\begin{bchapter}
\bauthor{\bsnm{Zhang}, \binits{Y.}},
\bauthor{\bsnm{Wang}, \binits{Z.}},
\bauthor{\bsnm{Wang}, \binits{D.}},
\bauthor{\bsnm{Xia}, \binits{G.}}:
\bctitle{Butter: A representation learning framework for bi-directional
  music-sentence retrieval and generation}.
In: \bbtitle{Proceedings of the 1st Workshop on Nlp for Music and Audio
  (nlp4musa)},
pp. \bfpage{54}--\blpage{58}
(\byear{2020})
\end{bchapter}
\endbibitem

%%% 26
\bibitem{DBLP:conf/ismir/0021WBD20}
\begin{bchapter}
\bauthor{\bsnm{Chen}, \binits{K.}},
\bauthor{\bsnm{Wang}, \binits{C.}},
\bauthor{\bsnm{Berg{-}Kirkpatrick}, \binits{T.}},
\bauthor{\bsnm{Dubnov}, \binits{S.}}:
\bctitle{Music sketchnet: Controllable music generation via factorized
  representations of pitch and rhythm}.
In: \beditor{\bsnm{Cumming}, \binits{J.}},
\beditor{\bsnm{Lee}, \binits{J.H.}},
\beditor{\bsnm{McFee}, \binits{B.}},
\beditor{\bsnm{Schedl}, \binits{M.}},
\beditor{\bsnm{Devaney}, \binits{J.}},
\beditor{\bsnm{McKay}, \binits{C.}},
\beditor{\bsnm{Zangerle}, \binits{E.}},
\beditor{\bparticle{de} \bsnm{Reuse}, \binits{T.}} (eds.)
\bbtitle{Proceedings of the 21th International Society for Music Information
  Retrieval Conference, {ISMIR} 2020, Montreal, Canada, October 11-16, 2020},
pp. \bfpage{77}--\blpage{84}
(\byear{2020}).
\burl{http://archives.ismir.net/ismir2020/paper/000146.pdf}
\end{bchapter}
\endbibitem

%%% 27
\bibitem{DBLP:conf/ismir/0008ZZJYXZ20}
\begin{bchapter}
\bauthor{\bsnm{Wang}, \binits{Z.}},
\bauthor{\bsnm{Zhang}, \binits{Y.}},
\bauthor{\bsnm{Zhang}, \binits{Y.}},
\bauthor{\bsnm{Jiang}, \binits{J.}},
\bauthor{\bsnm{Yang}, \binits{R.}},
\bauthor{\bsnm{Xia}, \binits{G.}},
\bauthor{\bsnm{Zhao}, \binits{J.}}:
\bctitle{{PIANOTREE} {VAE:} structured representation learning for polyphonic
  music}.
In: \beditor{\bsnm{Cumming}, \binits{J.}},
\beditor{\bsnm{Lee}, \binits{J.H.}},
\beditor{\bsnm{McFee}, \binits{B.}},
\beditor{\bsnm{Schedl}, \binits{M.}},
\beditor{\bsnm{Devaney}, \binits{J.}},
\beditor{\bsnm{McKay}, \binits{C.}},
\beditor{\bsnm{Zangerle}, \binits{E.}},
\beditor{\bparticle{de} \bsnm{Reuse}, \binits{T.}} (eds.)
\bbtitle{Proceedings of the 21th International Society for Music Information
  Retrieval Conference, {ISMIR} 2020, Montreal, Canada, October 11-16, 2020},
pp. \bfpage{368}--\blpage{375}
(\byear{2020}).
\burl{http://archives.ismir.net/ismir2020/paper/000096.pdf}
\end{bchapter}
\endbibitem

%%% 28
\bibitem{DBLP:conf/mm/DiJ0WZHLY21}
\begin{bchapter}
\bauthor{\bsnm{Di}, \binits{S.}},
\bauthor{\bsnm{Jiang}, \binits{Z.}},
\bauthor{\bsnm{Liu}, \binits{S.}},
\bauthor{\bsnm{Wang}, \binits{Z.}},
\bauthor{\bsnm{Zhu}, \binits{L.}},
\bauthor{\bsnm{He}, \binits{Z.}},
\bauthor{\bsnm{Liu}, \binits{H.}},
\bauthor{\bsnm{Yan}, \binits{S.}}:
\bctitle{Video background music generation with controllable music
  transformer}.
In: \beditor{\bsnm{Shen}, \binits{H.T.}},
\beditor{\bsnm{Zhuang}, \binits{Y.}},
\beditor{\bsnm{Smith}, \binits{J.R.}},
\beditor{\bsnm{Yang}, \binits{Y.}},
\beditor{\bsnm{Cesar}, \binits{P.}},
\beditor{\bsnm{Metze}, \binits{F.}},
\beditor{\bsnm{Prabhakaran}, \binits{B.}} (eds.)
\bbtitle{{MM} '21: {ACM} Multimedia Conference, Virtual Event, China, October
  20 - 24, 2021},
pp. \bfpage{2037}--\blpage{2045}.
\bpublisher{{ACM}}, \blocation{???}
(\byear{2021}).
doi:\doiurl{10.1145/3474085.3475195}.
\burl{https://doi.org/10.1145/3474085.3475195}
\end{bchapter}
\endbibitem

%%% 29
\bibitem{DBLP:journals/access/RhyuCKL22}
\begin{barticle}
\bauthor{\bsnm{Rhyu}, \binits{S.}},
\bauthor{\bsnm{Choi}, \binits{H.}},
\bauthor{\bsnm{Kim}, \binits{S.}},
\bauthor{\bsnm{Lee}, \binits{K.}}:
\batitle{Translating melody to chord: Structured and flexible harmonization of
  melody with transformer}.
\bjtitle{{IEEE} Access}
\bvolume{10},
\bfpage{28261}--\blpage{28273}
(\byear{2022}).
doi:\doiurl{10.1109/ACCESS.2022.3155467}
\end{barticle}
\endbibitem

%%% 30
\bibitem{harte2006detecting}
\begin{bchapter}
\bauthor{\bsnm{Harte}, \binits{C.}},
\bauthor{\bsnm{Sandler}, \binits{M.}},
\bauthor{\bsnm{Gasser}, \binits{M.}}:
\bctitle{Detecting harmonic change in musical audio}.
In: \bbtitle{Proceedings of the 1st ACM Workshop on Audio and Music Computing
  Multimedia},
pp. \bfpage{21}--\blpage{26}
(\byear{2006})
\end{bchapter}
\endbibitem

\end{thebibliography}

\newcommand{\BMCxmlcomment}[1]{}

\BMCxmlcomment{

<refgrp>

<bibl id="B1">
  <title><p>Residual Recurrent {CRNN} for End-to-End Optical Music Recognition
  on Monophonic Scores</p></title>
  <aug>
    <au><snm>Liu</snm><fnm>A</fnm></au>
    <au><snm>Zhang</snm><fnm>L</fnm></au>
    <au><snm>Mei</snm><fnm>Y</fnm></au>
    <au><snm>Han</snm><fnm>B</fnm></au>
    <au><snm>Cai</snm><fnm>Z</fnm></au>
    <au><snm>Zhu</snm><fnm>Z</fnm></au>
    <au><snm>Xiao</snm><fnm>J</fnm></au>
  </aug>
  <source>MMPT@ICMR2021: Proceedings of the 2021 Workshop on Multi-Modal
  Pre-Training for Multimedia Understanding, Taipei, Taiwan, August 21,
  2021</source>
  <publisher>{ACM}</publisher>
  <editor>Bei Liu and Jianlong Fu and Shizhe Chen and Qin Jin and Alexander G.
  Hauptmann and Yong Rui</editor>
  <pubdate>2021</pubdate>
  <fpage>23</fpage>
  <lpage>-27</lpage>
  <url>https://doi.org/10.1145/3463945.3469056</url>
</bibl>

<bibl id="B2">
  <title><p>Camera-PrIMuS: Neural End-to-End Optical Music Recognition on
  Realistic Monophonic Scores</p></title>
  <aug>
    <au><snm>Calvo{-}Zaragoza</snm><fnm>J</fnm></au>
    <au><snm>Rizo</snm><fnm>D</fnm></au>
  </aug>
  <source>Proceedings of the 19th International Society for Music Information
  Retrieval Conference, {ISMIR} 2018, Paris, France, September 23-27,
  2018</source>
  <editor>Emilia G{\'{o}}mez and Xiao Hu and Eric Humphrey and Emmanouil
  Benetos</editor>
  <pubdate>2018</pubdate>
  <fpage>248</fpage>
  <lpage>-255</lpage>
  <url>http://ismir2018.ircam.fr/doc/pdfs/33\_Paper.pdf</url>
</bibl>

<bibl id="B3">
  <title><p>Music Genre Recognition Using Deep Neural Networks and Transfer
  Learning</p></title>
  <aug>
    <au><snm>Ghosal</snm><fnm>D</fnm></au>
    <au><snm>Kolekar</snm><fnm>MH</fnm></au>
  </aug>
  <source>Interspeech 2018, 19th Annual Conference of the International Speech
  Communication Association, Hyderabad, India, 2-6 September 2018</source>
  <publisher>{ISCA}</publisher>
  <editor>B. Yegnanarayana</editor>
  <pubdate>2018</pubdate>
  <fpage>2087</fpage>
  <lpage>-2091</lpage>
  <url>https://doi.org/10.21437/Interspeech.2018-2045</url>
</bibl>

<bibl id="B4">
  <title><p>Genre Recognition from Symbolic Music with CNNs</p></title>
  <aug>
    <au><snm>Dervakos</snm><fnm>E</fnm></au>
    <au><snm>Kotsani</snm><fnm>N</fnm></au>
    <au><snm>Stamou</snm><fnm>G</fnm></au>
  </aug>
  <source>Artificial Intelligence in Music, Sound, Art and Design - 10th
  International Conference, EvoMUSART 2021, Held as Part of EvoStar 2021,
  Virtual Event, April 7-9, 2021, Proceedings</source>
  <publisher>Springer</publisher>
  <editor>Juan Romero and Tiago Martins and Nereida
  Rodr{\'{\i}}guez{-}Fern{\'{a}}ndez</editor>
  <series><title><p>Lecture Notes in Computer Science</p></title></series>
  <pubdate>2021</pubdate>
  <volume>12693</volume>
  <fpage>98</fpage>
  <lpage>-114</lpage>
  <url>https://doi.org/10.1007/978-3-030-72914-1\_7</url>
</bibl>

<bibl id="B5">
  <title><p>Deep Learning Techniques for Music Generation - {A}
  Survey</p></title>
  <aug>
    <au><snm>Briot</snm><fnm>J</fnm></au>
    <au><snm>Hadjeres</snm><fnm>G</fnm></au>
    <au><snm>Pachet</snm><fnm>F</fnm></au>
  </aug>
  <source>CoRR</source>
  <pubdate>2017</pubdate>
  <volume>abs/1709.01620</volume>
</bibl>

<bibl id="B6">
  <title><p>Some Reflections on the Potential and Limitations of Deep Learning
  for Automated Music Generation</p></title>
  <aug>
    <au><snm>Casini</snm><fnm>L</fnm></au>
    <au><snm>Marfia</snm><fnm>G</fnm></au>
    <au><snm>Roccetti</snm><fnm>M</fnm></au>
  </aug>
  <source>29th {IEEE} Annual International Symposium on Personal, Indoor and
  Mobile Radio Communications</source>
  <publisher>{IEEE}</publisher>
  <pubdate>2018</pubdate>
  <fpage>27</fpage>
  <lpage>-31</lpage>
</bibl>

<bibl id="B7">
  <title><p>A Functional Taxonomy of Music Generation Systems</p></title>
  <aug>
    <au><snm>Herremans</snm><fnm>D</fnm></au>
    <au><snm>Chuan</snm><fnm>C</fnm></au>
    <au><snm>Chew</snm><fnm>E</fnm></au>
  </aug>
  <source>{ACM} Comput. Surv.</source>
  <pubdate>2017</pubdate>
  <volume>50</volume>
  <issue>5</issue>
  <fpage>69:1</fpage>
  <lpage>-69:30</lpage>
</bibl>

<bibl id="B8">
  <title><p>Automatic melodic harmonization: An overview, challenges and future
  directions</p></title>
  <aug>
    <au><snm>Makris</snm><fnm>D</fnm></au>
    <au><snm>Kayrdis</snm><fnm>I</fnm></au>
    <au><snm>Sioutas</snm><fnm>S</fnm></au>
  </aug>
  <source>Trends in music information seeking, behavior, and retrieval for
  creativity</source>
  <publisher>IGI global</publisher>
  <pubdate>2016</pubdate>
  <fpage>146</fpage>
  <lpage>-165</lpage>
</bibl>

<bibl id="B9">
  <title><p>Melodic Skeleton: {A} Musical Feature for Automatic Melody
  Harmonization</p></title>
  <aug>
    <au><snm>Sun</snm><fnm>W</fnm></au>
    <au><snm>Wu</snm><fnm>J</fnm></au>
    <au><snm>Yuan</snm><fnm>S</fnm></au>
  </aug>
  <source>{IEEE} International Conference on Multimedia and Expo Workshops,
  {ICME} Workshops 2022, Taipei, Taiwan, July 18-22, 2022</source>
  <publisher>{IEEE}</publisher>
  <pubdate>2022</pubdate>
  <fpage>1</fpage>
  <lpage>-6</lpage>
</bibl>

<bibl id="B10">
  <title><p>Function- and Rhythm-Aware Melody Harmonization Based on
  Tree-Structured Parsing and Split-Merge Sampling of Chord
  Sequences</p></title>
  <aug>
    <au><snm>Tsushima</snm><fnm>H</fnm></au>
    <au><snm>Nakamura</snm><fnm>E</fnm></au>
    <au><snm>Itoyama</snm><fnm>K</fnm></au>
    <au><snm>Yoshii</snm><fnm>K</fnm></au>
  </aug>
  <source>Proceedings of the 18th International Society for Music Information
  Retrieval Conference, {ISMIR} 2017, Suzhou, China, October 23-27,
  2017</source>
  <pubdate>2017</pubdate>
  <fpage>502</fpage>
  <lpage>-508</lpage>
</bibl>

<bibl id="B11">
  <title><p>JamBot: Music Theory Aware Chord Based Generation of Polyphonic
  Music with LSTMs</p></title>
  <aug>
    <au><snm>Brunner</snm><fnm>G</fnm></au>
    <au><snm>Wang</snm><fnm>Y</fnm></au>
    <au><snm>Wattenhofer</snm><fnm>R</fnm></au>
    <au><snm>Wiesendanger</snm><fnm>J</fnm></au>
  </aug>
  <source>29th {IEEE} International Conference on Tools with Artificial
  Intelligence, {ICTAI} 2017, Boston, MA, USA, November 6-8, 2017</source>
  <publisher>{IEEE} Computer Society</publisher>
  <pubdate>2017</pubdate>
  <fpage>519</fpage>
  <lpage>-526</lpage>
</bibl>

<bibl id="B12">
  <title><p>Melody Harmonization Using Orderless Nade, Chord Balancing, and
  Blocked Gibbs Sampling</p></title>
  <aug>
    <au><snm>Sun</snm><fnm>C</fnm></au>
    <au><snm>Chen</snm><fnm>Y</fnm></au>
    <au><snm>Lee</snm><fnm>H</fnm></au>
    <au><snm>Chen</snm><fnm>Y</fnm></au>
    <au><snm>Wang</snm><fnm>H</fnm></au>
  </aug>
  <source>{IEEE} International Conference on Acoustics, Speech and Signal
  Processing, {ICASSP} 2021, Toronto, ON, Canada, June 6-11, 2021</source>
  <publisher>{IEEE}</publisher>
  <pubdate>2021</pubdate>
  <fpage>4145</fpage>
  <lpage>-4149</lpage>
</bibl>

<bibl id="B13">
  <title><p>Automatic Melody Harmonization with Triad Chords: {A} Comparative
  Study</p></title>
  <aug>
    <au><snm>Yeh</snm><fnm>Y</fnm></au>
    <au><snm>Hsiao</snm><fnm>W</fnm></au>
    <au><snm>Fukayama</snm><fnm>S</fnm></au>
    <au><snm>Kitahara</snm><fnm>T</fnm></au>
    <au><snm>Genchel</snm><fnm>B</fnm></au>
    <au><snm>Liu</snm><fnm>H</fnm></au>
    <au><snm>Dong</snm><fnm>H</fnm></au>
    <au><snm>Chen</snm><fnm>Y</fnm></au>
    <au><snm>Leong</snm><fnm>T</fnm></au>
    <au><snm>Yang</snm><fnm>Y</fnm></au>
  </aug>
  <source>CoRR</source>
  <pubdate>2020</pubdate>
  <volume>abs/2001.02360</volume>
</bibl>

<bibl id="B14">
  <title><p>SurpriseNet: Melody Harmonization Conditioning on User-controlled
  Surprise Contours</p></title>
  <aug>
    <au><snm>Chen</snm><fnm>Y</fnm></au>
    <au><snm>Lee</snm><fnm>H</fnm></au>
    <au><snm>Chen</snm><fnm>Y</fnm></au>
    <au><snm>Wang</snm><fnm>H</fnm></au>
  </aug>
  <source>Proceedings of the 22nd International Society for Music Information
  Retrieval Conference, {ISMIR} 2021, Online, November 7-12, 2021</source>
  <pubdate>2021</pubdate>
  <fpage>105</fpage>
  <lpage>-112</lpage>
</bibl>

<bibl id="B15">
  <title><p>Efficient and Training-Free Control of Language
  Generation</p></title>
  <aug>
    <au><snm>Wu</snm><fnm>S</fnm></au>
    <au><snm>Sun</snm><fnm>M</fnm></au>
  </aug>
  <source>CoRR</source>
  <pubdate>2022</pubdate>
  <volume>abs/2205.06036</volume>
</bibl>

<bibl id="B16">
  <title><p>Chord Generation from Symbolic Melody Using {BLSTM}
  Networks</p></title>
  <aug>
    <au><snm>Lim</snm><fnm>H</fnm></au>
    <au><snm>Rhyu</snm><fnm>S</fnm></au>
    <au><snm>Lee</snm><fnm>K</fnm></au>
  </aug>
  <source>Proceedings of the 18th International Society for Music Information
  Retrieval Conference, {ISMIR} 2017, Suzhou, China, October 23-27,
  2017</source>
  <pubdate>2017</pubdate>
  <fpage>621</fpage>
  <lpage>-627</lpage>
</bibl>

<bibl id="B17">
  <title><p>Automatic Stylistic Composition of Bach Chorales with Deep
  {LSTM}</p></title>
  <aug>
    <au><snm>Liang</snm><fnm>FT</fnm></au>
    <au><snm>Gotham</snm><fnm>M</fnm></au>
    <au><snm>Johnson</snm><fnm>M</fnm></au>
    <au><snm>Shotton</snm><fnm>J</fnm></au>
  </aug>
  <source>Proceedings of the 18th International Society for Music Information
  Retrieval Conference</source>
  <pubdate>2017</pubdate>
  <fpage>449</fpage>
  <lpage>-456</lpage>
</bibl>

<bibl id="B18">
  <title><p>DeepBach: a Steerable Model for Bach Chorales
  Generation</p></title>
  <aug>
    <au><snm>Hadjeres</snm><fnm>G</fnm></au>
    <au><snm>Pachet</snm><fnm>F</fnm></au>
    <au><snm>Nielsen</snm><fnm>F</fnm></au>
  </aug>
  <source>Proceedings of the 34th International Conference on Machine
  Learning</source>
  <publisher>{PMLR}</publisher>
  <pubdate>2017</pubdate>
  <volume>70</volume>
  <fpage>1362</fpage>
  <lpage>-1371</lpage>
</bibl>

<bibl id="B19">
  <title><p>The Bach Doodle: Approachable music composition with machine
  learning at scale</p></title>
  <aug>
    <au><snm>Huang</snm><fnm>CA</fnm></au>
    <au><snm>Hawthorne</snm><fnm>C</fnm></au>
    <au><snm>Roberts</snm><fnm>A</fnm></au>
    <au><snm>Dinculescu</snm><fnm>M</fnm></au>
    <au><snm>Wexler</snm><fnm>J</fnm></au>
    <au><snm>Hong</snm><fnm>L</fnm></au>
    <au><snm>Howcroft</snm><fnm>J</fnm></au>
  </aug>
  <source>CoRR</source>
  <pubdate>2019</pubdate>
  <volume>abs/1907.06637</volume>
  <url>http://arxiv.org/abs/1907.06637</url>
</bibl>

<bibl id="B20">
  <title><p>CLSTMS: A combination of two LSTM models to generate chords
  accompaniment for symbolic melody</p></title>
  <aug>
    <au><snm>Yang</snm><fnm>W</fnm></au>
    <au><snm>Sun</snm><fnm>P</fnm></au>
    <au><snm>Zhang</snm><fnm>Y</fnm></au>
    <au><snm>Zhang</snm><fnm>Y</fnm></au>
  </aug>
  <source>2019 International Conference on High Performance Big Data and
  Intelligent Systems (HPBD\&IS)</source>
  <pubdate>2019</pubdate>
  <fpage>176</fpage>
  <lpage>-180</lpage>
</bibl>

<bibl id="B21">
  <title><p>Music Harmony Generation, through Deep Learning and Using a
  Multi-Objective Evolutionary Algorithm</p></title>
  <aug>
    <au><snm>Majidi</snm><fnm>M</fnm></au>
    <au><snm>Toroghi</snm><fnm>RM</fnm></au>
  </aug>
  <source>CoRR</source>
  <pubdate>2021</pubdate>
  <volume>abs/2102.07960</volume>
</bibl>

<bibl id="B22">
  <title><p>Translating Melody to Chord: Structured and Flexible Harmonization
  of Melody With Transformer</p></title>
  <aug>
    <au><snm>Rhyu</snm><fnm>S</fnm></au>
    <au><snm>Choi</snm><fnm>H</fnm></au>
    <au><snm>Kim</snm><fnm>S</fnm></au>
    <au><snm>Lee</snm><fnm>K</fnm></au>
  </aug>
  <source>IEEE Access</source>
  <pubdate>2022</pubdate>
  <volume>10</volume>
  <fpage>28261</fpage>
  <lpage>28273</lpage>
</bibl>

<bibl id="B23">
  <title><p>Hierarchical Variational Autoencoders for Music</p></title>
  <source>Workshop on Machine Learning for Creativity and Design, NIPS</source>
  <editor>Adam Roberts and Jesse Engel and Douglas Eck</editor>
  <pubdate>2017</pubdate>
</bibl>

<bibl id="B24">
  <title><p>Learning Disentangled Representations of Timbre and Pitch for
  Musical Instrument Sounds Using Gaussian Mixture Variational
  Autoencoders</p></title>
  <aug>
    <au><snm>Luo</snm><fnm>Y</fnm></au>
    <au><snm>Agres</snm><fnm>K</fnm></au>
    <au><snm>Herremans</snm><fnm>D</fnm></au>
  </aug>
  <source>Proceedings of the 20th International Society for Music Information
  Retrieval Conference, {ISMIR} 2019, Delft, The Netherlands, November 4-8,
  2019</source>
  <editor>Arthur Flexer and Geoffroy Peeters and Juli{\'{a}}n Urbano and Anja
  Volk</editor>
  <pubdate>2019</pubdate>
  <fpage>746</fpage>
  <lpage>-753</lpage>
  <url>http://archives.ismir.net/ismir2019/paper/000091.pdf</url>
</bibl>

<bibl id="B25">
  <title><p>BUTTER: A representation learning framework for bi-directional
  music-sentence retrieval and generation</p></title>
  <aug>
    <au><snm>Zhang</snm><fnm>Y</fnm></au>
    <au><snm>Wang</snm><fnm>Z</fnm></au>
    <au><snm>Wang</snm><fnm>D</fnm></au>
    <au><snm>Xia</snm><fnm>G</fnm></au>
  </aug>
  <source>Proceedings of the 1st workshop on nlp for music and audio
  (nlp4musa)</source>
  <pubdate>2020</pubdate>
  <fpage>54</fpage>
  <lpage>-58</lpage>
</bibl>

<bibl id="B26">
  <title><p>Music SketchNet: Controllable Music Generation via Factorized
  Representations of Pitch and Rhythm</p></title>
  <aug>
    <au><snm>Chen</snm><fnm>K</fnm></au>
    <au><snm>Wang</snm><fnm>C</fnm></au>
    <au><snm>Berg{-}Kirkpatrick</snm><fnm>T</fnm></au>
    <au><snm>Dubnov</snm><fnm>S</fnm></au>
  </aug>
  <source>Proceedings of the 21th International Society for Music Information
  Retrieval Conference, {ISMIR} 2020, Montreal, Canada, October 11-16,
  2020</source>
  <editor>Julie Cumming and Jin Ha Lee and Brian McFee and Markus Schedl and
  Johanna Devaney and Cory McKay and Eva Zangerle and Timothy de Reuse</editor>
  <pubdate>2020</pubdate>
  <fpage>77</fpage>
  <lpage>-84</lpage>
  <url>http://archives.ismir.net/ismir2020/paper/000146.pdf</url>
</bibl>

<bibl id="B27">
  <title><p>{PIANOTREE} {VAE:} Structured Representation Learning for
  Polyphonic Music</p></title>
  <aug>
    <au><snm>Wang</snm><fnm>Z</fnm></au>
    <au><snm>Zhang</snm><fnm>Y</fnm></au>
    <au><snm>Zhang</snm><fnm>Y</fnm></au>
    <au><snm>Jiang</snm><fnm>J</fnm></au>
    <au><snm>Yang</snm><fnm>R</fnm></au>
    <au><snm>Xia</snm><fnm>G</fnm></au>
    <au><snm>Zhao</snm><fnm>J</fnm></au>
  </aug>
  <source>Proceedings of the 21th International Society for Music Information
  Retrieval Conference, {ISMIR} 2020, Montreal, Canada, October 11-16,
  2020</source>
  <editor>Julie Cumming and Jin Ha Lee and Brian McFee and Markus Schedl and
  Johanna Devaney and Cory McKay and Eva Zangerle and Timothy de Reuse</editor>
  <pubdate>2020</pubdate>
  <fpage>368</fpage>
  <lpage>-375</lpage>
  <url>http://archives.ismir.net/ismir2020/paper/000096.pdf</url>
</bibl>

<bibl id="B28">
  <title><p>Video Background Music Generation with Controllable Music
  Transformer</p></title>
  <aug>
    <au><snm>Di</snm><fnm>S</fnm></au>
    <au><snm>Jiang</snm><fnm>Z</fnm></au>
    <au><snm>Liu</snm><fnm>S</fnm></au>
    <au><snm>Wang</snm><fnm>Z</fnm></au>
    <au><snm>Zhu</snm><fnm>L</fnm></au>
    <au><snm>He</snm><fnm>Z</fnm></au>
    <au><snm>Liu</snm><fnm>H</fnm></au>
    <au><snm>Yan</snm><fnm>S</fnm></au>
  </aug>
  <source>{MM} '21: {ACM} Multimedia Conference, Virtual Event, China, October
  20 - 24, 2021</source>
  <publisher>{ACM}</publisher>
  <editor>Heng Tao Shen and Yueting Zhuang and John R. Smith and Yang Yang and
  Pablo Cesar and Florian Metze and Balakrishnan Prabhakaran</editor>
  <pubdate>2021</pubdate>
  <fpage>2037</fpage>
  <lpage>-2045</lpage>
  <url>https://doi.org/10.1145/3474085.3475195</url>
</bibl>

<bibl id="B29">
  <title><p>Translating Melody to Chord: Structured and Flexible Harmonization
  of Melody With Transformer</p></title>
  <aug>
    <au><snm>Rhyu</snm><fnm>S</fnm></au>
    <au><snm>Choi</snm><fnm>H</fnm></au>
    <au><snm>Kim</snm><fnm>S</fnm></au>
    <au><snm>Lee</snm><fnm>K</fnm></au>
  </aug>
  <source>{IEEE} Access</source>
  <pubdate>2022</pubdate>
  <volume>10</volume>
  <fpage>28261</fpage>
  <lpage>-28273</lpage>
</bibl>

<bibl id="B30">
  <title><p>Detecting harmonic change in musical audio</p></title>
  <aug>
    <au><snm>Harte</snm><fnm>C</fnm></au>
    <au><snm>Sandler</snm><fnm>M</fnm></au>
    <au><snm>Gasser</snm><fnm>M</fnm></au>
  </aug>
  <source>Proceedings of the 1st ACM workshop on Audio and music computing
  multimedia</source>
  <pubdate>2006</pubdate>
  <fpage>21</fpage>
  <lpage>-26</lpage>
</bibl>

</refgrp>
} % end of \BMCxmlcomment

\end{backmatter}
\end{document}